\documentclass[12pt]{article}
\usepackage{a4wide}
\usepackage{amssymb}
\usepackage{amsmath}
\usepackage{graphicx}
\usepackage{mathdots}
\usepackage{slashed}
\usepackage{verbatim}
\usepackage{caption}
\usepackage{subcaption}

 \def\makeatletter{\catcode`\@=11}
 \makeatletter
 \def\mathbox#1{\hbox{$\m@th#1$}}%
\def\math@ccstyles#1#2#3#4#5#6#7{{\leavevmode
      \setbox0\mathbox{#6#7}%
      \setbox2\mathbox{#4#5}%
      \dimen@ #3%
      \baselineskip\z@\lineskiplimit#1\lineskip\z@
      \vbox{\ialign{##\crcr
             \hfil \kern #2\box2 \hfil\crcr
             \noalign{\kern\dimen@}%
             \hfil\box0\hfil\crcr}}}}
\def\mathaccstyles{\math@ccstyles\maxdimen}
\def\maththroughstyles{\math@ccstyles{-\maxdimen}}
\def\unity%
 {\maththroughstyles{.45\ht0}\z@\displaystyle {\mathchar"006C}\displaystyle 1}
 
\usepackage{array}
\usepackage{amsthm}

 \numberwithin{equation}{section}

\begin{document}

\mbox{}
\vspace{0truecm}
\linespread{1.1}


%
%
%
%
%
%

\centerline{\Large \bf Defects in  scalar field theories, RG flows}

\bigskip

\centerline{\Large \bf and Dimensional Disentangling}


\vspace{.4cm}

 \centerline{\LARGE \bf }

\vspace{1.5truecm}

\centerline{
    { \bf D. Rodriguez-Gomez${}^{a,b}$} \footnote{d.rodriguez.gomez@uniovi.es}
   {\bf and}
    { \bf J. G. Russo ${}^{c,d}$} \footnote{jorge.russo@icrea.cat}}

\vspace{1cm}
\centerline{{\it ${}^a$ Department of Physics, Universidad de Oviedo}} \centerline{{\it C/ Federico Garc\'ia Lorca  18, 33007  Oviedo, Spain}}
\medskip
\centerline{{\it ${}^b$  Instituto Universitario de Ciencias y Tecnolog\'ias Espaciales de Asturias (ICTEA)}}\centerline{{\it C/~de la Independencia 13, 33004 Oviedo, Spain.}}
\medskip
\centerline{{\it ${}^c$ Instituci\'o Catalana de Recerca i Estudis Avan\c{c}ats (ICREA)}} \centerline{{\it Pg.~Lluis Companys, 23, 08010 Barcelona, Spain}}
\medskip
\centerline{{\it ${}^d$ Departament de F\' \i sica Cu\' antica i Astrof\'\i sica and Institut de Ci\`encies del Cosmos}} \centerline{{\it Universitat de Barcelona, Mart\'i Franqu\`es, 1, 08028
Barcelona, Spain }}
\vspace{1cm}

\centerline{\bf ABSTRACT}
\medskip

We consider defect operators in scalar field theories in  dimensions $d=4-\epsilon $ and $d=6-\epsilon$ with self-interactions given by a general marginal potential. In  a double scaling limit, where the bulk couplings go to zero and the defect couplings go to infinity, the bulk theory becomes classical and the quantum defect theory can be solved order by order in perturbation theory. We compute the defect $\beta $ functions to two loops and study the Renormalization Group flows. The defect fixed points can move and merge, leading to fixed point annihilation; and they exhibit a remarkable factorization property where the $\epsilon$-dependence gets disentangled from the coupling dependence.

\noindent 

\newpage

\tableofcontents

\section{Introduction}

Besides the familiar local operators supported on points, a Quantum Field Theory can contain extended operators supported on more general manifolds. These provide the key to the modern notion of generalized symmetry, which is based on extended operators that depend only topologically on the manifold on which they are defined. In other instances, defects provide a new window into the structure of Quantum Field  Theory as they can probe aspects of Renormalization Group (RG) flows or detect new central charges associated to a given theory. In some physical systems, extended operators can describe impurities. Thus the study of defects may give important hints
on long-standing questions in condensed matter regarding the ground state, large distance physics and RG flows in the presence of impurities.

In this paper, we  investigate  defects in  self-interacting scalar field theories with general classically marginal potentials.\footnote{By ``classically marginal" we mean that they are  marginal potentials in the classical theory with $\epsilon=0$.}
The coupling of the the bulk theory to the extended operator triggers an RG flow. Our aim is to study this RG flow and the possible fixed points on which the defect theory may end. A particularly interesting case is that in which the bulk theory is itself a Conformal Field Theory (CFT), when the critical defect defines a so-called defect Conformal Field Theory (dCFT) (see  \cite{Billo:2016cpy} for general properties of dCFT's). 
General properties of line defect RG flows in arbitrary dimensions have been uncovered in \cite{Cuomo:2021rkm}, extending the 2d results of \cite{Affleck:1991tk}.

The simplest line defect one may imagine is a trivial line defect (defined by the insertion of the identity operator on a line, and whose only excitations are simply the bulk fields restricted to the line), which, as such, is not very interesting. However, this defect admits exactly one 
deformation consisting on a linear combination of the scalars of the theory. Our interest in this paper is to explore the properties  of the theory in the presence of this line defect. This system has been studied in the past in \cite{Allais:2014fqa} and revived more recently in \cite{Cuomo:2021kfm,Cuomo:2022xgw,Popov:2022nfq}.

Stated in full generality, the problem of studying RG flows on line defects in general scalar field theories (yet with classically marginal potentials) is very hard. However a particular double-scaling limit of the defect and bulk couplings was considered in \cite{Rodriguez-Gomez:2022gbz},   leading to dramatic simplifications. 
This is similar in spirit to the large charge limit considered in the literature in \textit{e.g.} \cite{Bourget:2018obm,Arias-Tamargo:2019xld,Watanabe:2019pdh,Badel:2019oxl} (a  review and further references can be found in \cite{Gaume:2020bmp}), with direct application to boundary CFT's in \cite{Cuomo:2021cnb}. In the  context of Wilson lines in gauge theories, similar limits have been  recently considered in \cite{Beccaria:2017rbe,Beccaria:2018ocq,Beccaria:2021rmj,Beccaria:2022bcr,Giombi:2021zfb,Giombi:2022anm,Cuomo:2022xgw, Rodriguez-Gomez:2022xwm}. It is worth noting that the double scaling limit considered in this paper may not only be a simplifying regime allowing for detailed studies, as it may also have phenomenological implications. In particular, as argued in \cite{Cuomo:2022xgw} in the case of the $O(3$) model in $d=4-\epsilon$ dimensions,  this limit provides an effective description of a large spin impurity in an antiferromagnet (see also \cite{Allais:2014fqa,Cuomo:2021kfm}), known as pinning field defect.

The double-scaling limit has the effect of suppressing the Feynman loop diagrams  involving bulk couplings. Thanks to  this dramatic simplification, we will be able to provide exact formulas for the $\beta$ functions up to two loops in the defect couplings for  theories with arbitrary potential. The RG flow resulting from these $\beta$ functions turns out to be a gradient flow, as the $\beta$ functions are the gradient of a scalar function $\mathcal{H}$. 
 
We find a rich structure of fixed points, which merge and disappear though fixed point annihilation as the ratio of bulk couplings are gradually changed. 
Strikingly, to 2 loops order, the $\epsilon$ dependence in the location of the fixed points factorizes (this means that the $\epsilon$ dependence gets disentangled --\textit{i.e.} separated-- from the coupling dependence). A close examination of the higher-order corrections
to the $\beta$ functions gives a strong hint that this property should hold to all orders. This leads to an exact determination of the    fixed points and of the structure of the RG flows to all loop order.

\section{Defects in $d$-dimensional scalar field theories}

We consider a generic theory with $N$ scalar fields $\varphi_i$ in $d$ dimensions with a (classically) marginal potential. The action is

\begin{equation}
\mathcal{S}=\int d^d x\left( \frac{1}{2}\big(\partial\varphi_i\big)^2+V(\varphi_i)\right)\,.
\end{equation}
Here $V(\varphi_i)$ is a degree $n$ polynomial in the scalar fields $\varphi_i$ with couplings denoted generically by $\mathfrak{g}_\alpha$, where  $n=4$ in the $d=4-\epsilon$ theory and $n=3$ in the $d=6-\epsilon $ theory.

We now consider, in this theory, a trivial defect supported on a line in the 4d theory and on a two-dimensional plane in the $6d$ theory (that is, denoting the worldvolume dimension by $d_{\rm wv}$, in  4d, $d_{\rm wv}=1$ and in 6d, $d_{\rm wv}=2$). To introduce the defect,
the theory is deformed by an operator, which is a linear combination of the scalar fields in the theory. The deformed action is

\begin{equation}
\mathcal{S}\rightarrow \mathcal{S} = \int d^d x\left( \frac{1}{2} \big(\partial\varphi_i\big)^2+V(\varphi_i)+\mathfrak{h}_i\,\varphi_i\,\delta_T(\vec{x})\right)\,,
\end{equation}
where $\delta_T(\vec{x})$ is a delta function in the transverse space to the defect (which is $d-d_{\rm wv}$ dimensional). We will study the RG flow induced by the defect in a limit where
the coefficients $\mathfrak{h}_i$ are large. It is convenient to introduce a parameter $q$ to organize the limit, so that

\begin{equation}
(\mathfrak{g}_\alpha,\,\mathfrak{h}_i)=(\frac{g_\alpha}{q^{\frac{n-2}{2}}},\,\sqrt{q}\,\nu_i)\,.
\end{equation}
Upon writing $\varphi_i=\sqrt{q}\,\phi_i$, we have

\begin{equation}
\label{esses}
\mathcal{S} =q\,S\,,\qquad S= \int d^d x\left( \frac{1}{2}\big(\partial\phi_i\big)^2+V(\phi_i)+\nu_i\,\phi_i\,\delta_T(\vec{x})\right)\,.
\end{equation}
Let us now consider the limit $q\rightarrow \infty$ with fixed $g_\alpha,\,\nu_i$. 
In $d=4-\epsilon$, this corresponds to a double scaling limit  $\mathfrak{h}_i\to\infty$, $\mathfrak{g}_\alpha\to 0$, with fixed $\mathfrak{h}_i^2\mathfrak{g}_\alpha $. In $d=6-\epsilon $, the combination
$\mathfrak{h}_i\mathfrak{g}_\alpha$ is fixed.

In this limit we can use the saddle point method. Clearly, the limit corresponds to the classical
limit of the theory (with $q$ playing the role of $1/\hbar$) and therefore all bulk loop diagrams will be suppressed
(see section 3 for details). In turn, the defect theory will receive quantum corrections. 


The classical equations of motion are

\begin{equation}
\partial^2\phi_i=\frac{\delta V}{\delta\phi_i}+\nu_i\,\delta_T(\vec{x})\,.
\end{equation}
In order to do perturbation theory in the $g_\alpha $'s, we write $\phi_i=\phi_i^{(0)}+\phi_i^{(1)}+\phi_i^{(2)}\cdots$ where $\phi_i^{(p)}$ is of order $p$ in the $g_\alpha $'s. Substituting the perturbative expansion into the equations of motion and expanding to quadratic order in the $g_\alpha $'s , we obtain

\begin{equation}
\partial^2\phi_i^{(0)}+\partial^2\phi_i^{(1)}+\partial^2\phi_i^{(2)}=V_i+V_{ij}\phi_j^{(1)}+\nu_i\,\delta_T(\vec{x})\,.
\end{equation}
where we have used the obvious shorthand notation

\begin{equation}
V_i=\frac{\delta V}{\delta \phi_i}\Big|_{\phi_i^{(0)}}\,,\qquad V_{ij}=\frac{\delta^2V}{\delta\phi_i\delta\phi_j}\Big|_{\phi_i^{(0)}}\,.
\end{equation}
Solving order by order, we are led to the equations, 

\begin{equation}
\partial^2\phi_i^{(0)}=\nu_i\,\delta_T(\vec{x})\,,\qquad \partial^2\phi_i^{(1)}=V_i\,, \qquad \partial^2\phi_i^{(2)}=V_{ij}\,\phi_j^{(1)}\, .
\end{equation}
These equations are easily solved by using the Green's function. One has

\begin{eqnarray}
&&\phi_i^{(0)}=-\nu_i\int d^dz\,G(x-z)\,\delta_T(\vec{z})\,,\qquad \phi_i^{(1)}=-\int d^dz\,G(x-z)\,V_i(z)\,
\nonumber\\
&&\phi_i^{(2)}=-\int d^dz\,G(x-z) \,V_{ij}(z)\,\phi_j^{(1)}(z)\,.
\end{eqnarray}

Let us now evaluate the on-shell action. Using the equations of motion, the action takes the form

\begin{equation}
\label{SosEOM}
S_{\rm os}=\int d^d x\left(V-\frac{1}{2}\phi_i\,V_i+\frac{1}{2}\,\nu_i\,\phi_i\,\delta_T(\vec{x})\right)\,.
\end{equation}
Substituting the perturbative expansion, to quadratic order in the $g_\alpha $'s, we find

\begin{eqnarray}
S_{\rm os}&=&\int d^d x\, \Big( -\frac{1}{2}\,\phi^{(0)}_i\,V_i-\frac{1}{2}\,V_{ij}\, \phi_i^{(0)}\,\phi_j^{(1)}+\frac{1}{2}\,V_i\,\phi_i^{(1)}+V
\nonumber\\
&+&\frac{1}{2}\,\nu_i\,\phi^{(0)}_i\,\delta_T(\vec{x})+\frac{1}{2}\,\nu_i\,\phi^{(1)}_i\,\delta_T(\vec{x})+\frac{1}{2}\,\nu_i\,\phi^{(2)}_i\,\delta_T(\vec{x})\Big)\,.
\end{eqnarray}
From the explicit form of the perturbative solution one can easily show that

\begin{equation}
\int d^dx\,\frac{1}{2}\,\nu_i\,\phi^{(1)}_i(x)\,\delta_T(\vec{x})=\int d^dx\,\frac{1}{2}\, V_i(x)\, \phi_i^{(0)}(x)\,,
\end{equation}
and

\begin{equation}
\int d^dx\, \frac{1}{2}\,\nu_i\,\phi^{(2)}_i(x)\,\delta_T(\vec{x}) =\int d^dx\,\frac{1}{2}\, V_{ij}(x)\,\phi_i^{(0)}(x)\,\phi_j^{(1)}(x)\, .
\end{equation}
Hence

\begin{equation}
S_{\rm os}=\int d^d x\, \left(V+\frac{1}{2}V_i\,\phi_i^{(1)}+\frac{1}{2}\,\nu_i\,\phi^{(0)}_i\,\delta_T(\vec{x})\right)\,.
\end{equation}

In order to further proceed, we note that $\phi_i^{(0)}=\frac{\nu_i}{\nu_1}\,\phi_1^{(0)}$. Moreover, since $V$, $V_i$ are homogeneous functions of degree $n$ and $n-1$ respectively, it follows that

\begin{equation}
V=V(\phi_i^{(0)})=V(\nu_i)\,\frac{(\phi_1^{(0)})^n}{\nu_1^n}\,,\qquad V_i=\frac{\delta V}{\delta \phi_i}(\phi_i^{(0)})=V_i(\nu_i)\,\frac{(\phi_1^{(0)})^{n-1}}{\nu_1^{n-1}}\,.
\end{equation}
Therefore

\begin{equation}
\phi_i^{(1)}=-\frac{V_i(\nu_i)}{\nu_1^{n-1}}\,\int d^dy\,G(x-y)\,(\phi_1^{(0)})^{n-1}(y)\,.
\end{equation}
Then  

\begin{eqnarray}
S_{\rm os}&=&\frac{V}{\nu_1^n}\int d^dx\,(\phi_1^{(0)})^n(x)-\frac{V_i^2}{2\nu_1^{2(n-1)}}\,\int d^dx d^dy\,G(x-y)\,(\phi_1^{(0)})^{n-1}(x)(\phi_1^{(0)})^{n-1}(y)
\nonumber\\
&+&\frac{1}{2}\,\nu_i\,\int d^dx\,\phi^{(0)}_i\,\delta_T(\vec{x})\,,
\end{eqnarray}
where $V_i^2=\sum_i V_iV_i$ and in $V$, $V_i$ the $\phi_i$ are replaced by $\nu_i$. Using now the explicit form of $\phi_i^{(0)}$, we get

\begin{equation}
\label{Sos}
S_{\rm os}=-\frac{1}{2}\,\nu_i^2\,I_0+(-1)^n\,V(\nu_i)\,I_1-\frac{V_i^2}{2}\,I_2\,,
\end{equation}
where

\begin{eqnarray}
&&I_0=\int d^dz_1 \int d^dz_2\, G(z_2-z_2)\,\delta_T(\vec{z}_1) \delta_T(\vec{z}_2)\,,\\
&&I_1=\int d^dx\,\int d^dz_1\cdots \int d^dz_n \,G(x-z_1)\cdots G(x-z_n)\,\delta_T(\vec{z}_1)\cdots \delta_T(\vec{z}_n)\,,\\
&&I_2=\int d^dx \int d^dy \int d^du_1\cdots \int d^du_{n-1}\int d^dv_1\cdots \int d^dv_{n-1} G(x-y)\\ \nonumber && G(x-u_1)\cdots G(x-u_{n-1})\,G(y-v_1)\cdots G(y-v_{n-1})\,\delta_T(\vec{u}_1)\cdots \delta_T(\vec{u}_{n-1})\,\delta_T(\vec{v}_1)\cdots \delta_T(\vec{v}_{n-1})\,.
\end{eqnarray}

The factor $(-1)^n$ in the second term in \eqref{Sos} would give a $(-1)$ factor in the
$d=6-\epsilon$ theory. However, since in this case the potential is odd, we can redefine the scalar fields $\phi_i\to-\phi_i$ and set this factor to $+1$, so in what follows this factor will be removed.

\smallskip

Fourier-transforming one has 

\begin{equation}
\label{I's}
I_0= \mathcal{V}\,\int \frac{d\vec{p}_T}{(2\pi)^{d_{T}}}\frac{1}{|\vec{p}_T|^2}\,,\qquad I_1=\mathcal{V}\,\int \frac{d\vec{p}_T}{(2\pi)^{d_{T}}}\frac{1}{|\vec{p}_T|^2}\,\mathcal{I}_1\,,\qquad I_2=\mathcal{V}\,\int \frac{d\vec{p}_T}{(2\pi)^{d_{T}}}\frac{1}{|\vec{p}_T|^2}\,\mathcal{I}_2\,,
\end{equation}
where $\mathcal{V}$ is the regularized volume of the defect worldvolume, $d_T=d-d_{\rm wv}$ and $d\vec{p}_T=d^{d_T}p$,  represents the volume element in momentum space transverse to the defect and

\begin{equation}
\label{calI1}
\mathcal{I}_1=\int  \frac{d\vec{k}_T^1}{(2\pi)^{d_T}}\cdots \int  \frac{d\vec{k}_T^{n-2}}{(2\pi)^{d_T}}   \frac{1}{(\vec{k}_T^1)^2\cdots(\vec{k}^{n-2}_T)^2 \, (\vec{p}_T-\vec{k}_T^1-\cdots-\vec{k}_T^{n-2})^2} \,,
\end{equation}
and

\begin{eqnarray}
\label{calI2}
&&\mathcal{I}_2= \int  \frac{d\vec{k}_T^1}{(2\pi)^{d_T}}\cdots  \int \frac{d\vec{k}_T^{n-1}}{(2\pi)^{d_T}} \int  \frac{d\vec{q}_T^{\ 1}}{(2\pi)^{d_T}}\cdots  \int  \frac{d\vec{q}_T^{\ n-3}}{(2\pi)^{d_T}} \frac{1}{(\vec{k}_T^1)^2\cdots \, (\vec{k}_T^{n-1})^2}\,\\ \nonumber &&  \frac{1}{(\vec{k}_T^{n-1}-\vec{k}_T^1-\cdots-\vec{k}_T^{n-2})^2\,(\vec{q}_T^{\ 1})^2\cdots(\vec{q}_T^{\ n-3})^2 (\vec{k}_T^{n-1}+\vec{q}^{\ 1}_T+\cdots+\vec{q}^{\ n-3}_T-\vec{p}_T)^2} 
\,.
\end{eqnarray}
The integrals \eqref{calI1} and \eqref{calI2} are computed in the appendix. We obtain

\begin{equation}
\label{iuno}
\mathcal{I}_1=\frac{\Omega}{\epsilon}-2\,c\,\Omega\,\log|\vec{p}_T| + \epsilon\,A+\epsilon\,2\,c^2\,\Omega\,(\log|\vec{p}_T|)^2+O(\epsilon^2) \, , 
\end{equation}
and

\begin{equation}
\label{idose}
\mathcal{I}_2=\frac{\Omega^2}{2\epsilon^2}+\frac{c\,\Omega^2\,(c-2\,\log|\vec{p}_T|)}{\epsilon}+\Big(C-4\,c^3\,\Omega^2\,\log|\vec{p}_T|+4\,c^2\,\Omega^2\,(\log|\vec{p}_T|)^2\Big)+O(\epsilon)\,,
\end{equation}
where $A,\, C$ are (scheme-dependent) numerical constants. The table \eqref{tablad} summarizes the values of the different
constants in the four-dimensional and six-dimensional theories.

\begin{equation}
\label{tablad}
    \begin{array}{|c|c|c|c|c|c|} \hline
    d & \Omega & n & c & A & C \\ \hline
    4 & \frac{1}{32\pi^2}&4&1 & \frac{9\,\Omega}{2}-\frac{7}{768} & \frac{19\,\Omega^2}{2} -\frac{11\,\Omega}{768}\\ \hline
    6 & \frac{1}{8\pi^2}&3& \frac{1}{2} & \frac{\Omega}{2} -\frac{1}{384}& \frac{7\,\Omega^2}{8}-\frac{\Omega}{384}\\ \hline
    \end{array}
\end{equation}

\smallskip

\noindent Note that in $\mathcal{I}_1$ we have kept a term of  order $\epsilon$, whose relevance will become clear below.

Coming back to our computation, the on-shell action takes the form, $S_{\rm os}={\cal V}\, s_{\rm os}$, with

\begin{eqnarray}
\label{accionfinal}
s_{\rm os}&=&\Big(-\frac{\nu_i^2}{2}+\frac{2\,\Omega\,V-c^2\,\Omega^2\,V_i^2}{2\,\epsilon}-\frac{\Omega^2\,V_i^2}{4\,\epsilon^2}\Big)\int \frac{d\vec{p}_T}{(2\pi)^{d_T}}\frac{1}{|\vec{p}_T|^2}\\ \nonumber && -\Big(  2\,c\,\Omega\,V -\frac{c\,\Omega^2\,V_i^2}{\epsilon} \Big)\int \frac{d\vec{p}_T}{(2\pi)^{d_T}}\frac{\log|\vec{p}_T|}{|\vec{p}_T|^2} \\ \nonumber &&-\frac{V_i^2}{2}\,\int \frac{d\vec{p}_T}{(2\pi)^{d_T}}\,\frac{1}{|\vec{p}_T|^2}\,\Big(C-4\,c^3\,\Omega^2\,\log|\vec{p}_T|+4\,c^2\,\Omega^2\,(\log|\vec{p}_T|)^2\Big)
\\ \nonumber &&
+V\,\epsilon\,\int \frac{d\vec{p}_T}{(2\pi)^{d_T}}\,\frac{1}{p_T^2}\,\Big(A+2\,c^2\,\Omega\,(\log|\vec{p}_T|)^2\Big)\,.
\end{eqnarray}
The factors of $\Omega $ in \eqref{accionfinal} could be removed by a rescaling $\Omega g_\alpha\to g_\alpha$ along with a redefinition of the numerical constants $C$, $A$. However, we find more convenient to keep the explicit dependence on $\Omega$ in all formulas.

\section{Renormalization and $\beta$ functions}

The defect breaks translational invariance in the transverse directions. As a consequence, one-point functions in the bulk can be non-vanishing.\footnote{
For critical defects in bulk conformal theories, the $SO(d_{\rm wv}+1,1)\times SO(d-d_{\rm wv})$ conformal group does not imply the vanishing of the one-point function but it places other constraints on the structure of correlation functions. See \textit{e.g.} \cite{Billo:2016cpy} for a detailed discussion.} 
%
%
%
The integrated one-point function 
of the scalar in the presence of the defect can be computed by differentiating with respect to the defect coupling as 

\begin{equation}
\frac{1}{Z}\,\frac{\delta Z}{\delta\mathfrak{h}_i}=-\langle \int d^dx\,\varphi_i\,\delta_T(\vec{x})\rangle\qquad \longrightarrow \qquad \frac{1}{Z}\,\frac{\delta Z}{\delta\nu_i}=-q\,\langle \int d^dx\, \phi_i\,\delta_T(\vec{x})\rangle\,.
\end{equation}
In appendix A we review some basic facts about one-point functions in defect theories. Related calculations of the 1-point function can be found in \cite{Allais:2014fqa,Cuomo:2021kfm,Cuomo:2022xgw,Popov:2022nfq}.
In the large $q$ limit, this VEV is computed by

\begin{equation}
\label{VEV}
\langle \int d^dx\, \phi_i\,\delta_T(\vec{x})\rangle=- {\cal V}\,\partial_i s_{\rm os}\,,
\end{equation}
where $\partial_i$ is the derivative with respect to $\nu_i$. Using \eqref{accionfinal}, we obtain

\begin{eqnarray}
\label{dS}
\partial_is_{\rm os}&=&\Big(-\nu_i+\frac{ \Omega\,V_i-c^2\,\Omega^2\,V_j\,V_{ij}}{\epsilon}-\frac{\Omega^2\,V_j\,V_{ij}}{2\,\epsilon^2}\Big)\int \frac{d\vec{p}_T}{(2\pi)^{d_T}}\frac{1}{|\vec{p}_T|^2}\\ \nonumber && -2\,c\,\Big(  \Omega\,V_i -\frac{\Omega^2\,V_j\,V_{ij}}{\epsilon} \Big)\int \frac{d\vec{p}_T}{(2\pi)^{d_T}}\frac{\log|\vec{p}_T|}{|\vec{p}_T|^2} \\ \nonumber &&-V_j\,V_{ij}\,\int \frac{d\vec{p}_T}{(2\pi)^{d_T}}\,\frac{1}{|\vec{p}_T|^2}\,\Big(C-4\,c^3\,\Omega^2\,\log|\vec{p}_T|+4\,c^2\,\Omega^2\,(\log|\vec{p}_T|)^2\Big)
\\ \nonumber &&
+V_i\,\epsilon\,\int \frac{d\vec{p}_T}{(2\pi)^{d_T}}\,\frac{1}{p_T^2}\,\Big(A+2\,c^2\,\Omega\,(\log|\vec{p}_T|)^2\Big)\,.
\end{eqnarray}
This represents  the Fourier-transformed one-point function in the bulk (see appendix \ref{1pointappendix} for further discussions).

\begin{figure}[h!]
    \centering
    \includegraphics[scale=.25]{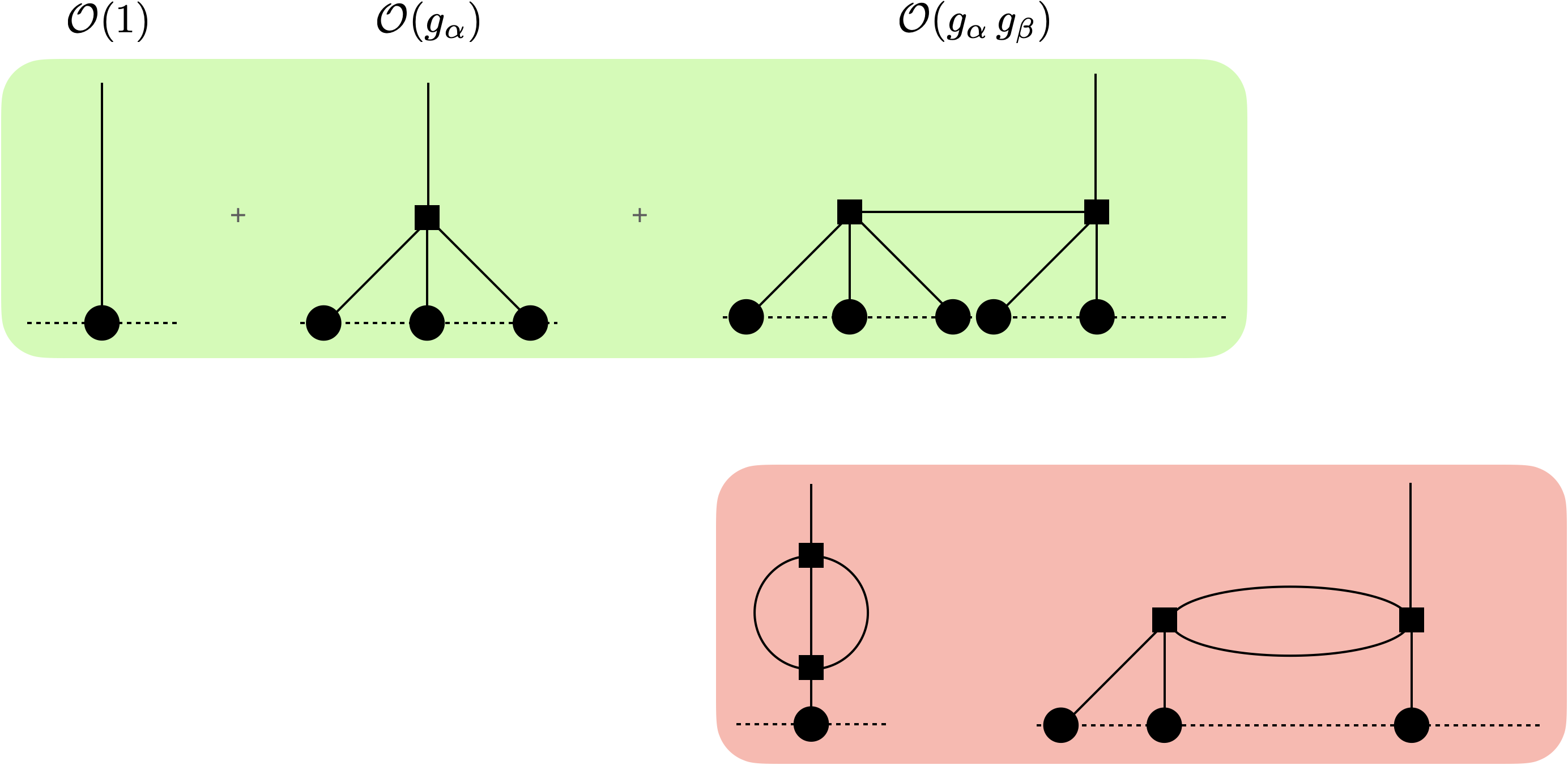}
    \caption{Diagrams  contributing to the one-point function of the scalar. Squares correspond to the bulk coupling; circles to defect couplings. In the double scaling limit only the diagrams with green background contribute, while those in red are suppressed.}
    \label{fig:diagrams}
\end{figure}

The content of \eqref{dS} as well as the simplifications that occur in this limit can be  understood diagrammatically from fig. \ref{fig:diagrams}, where
we list all diagrams  contributing to the one-point function. 
The vertices can be read from \eqref{Sos}, by computing $\partial_i S_{\rm os}$. The second and third diagrams in the first line (green background) give contributions proportional to the integrals $I_1$ and $I_2$, respectively. The diagrams in the second line in fig. \ref{fig:diagrams} (in   red background) are  $\mathcal{O}(g_{\alpha}g_{\beta})$ terms which are not contributing to  \eqref{dS}. They represent loops in the bulk couplings $g_\alpha$, which are suppressed in the
large $q$ (or classical bulk) limit considered here.\footnote{The explicit $q$-dependence is as follows.
The diagrams in green are proportional to $\mathfrak{h}_i\,(\mathfrak{h_j}\mathfrak{h_k}\mathfrak{g}_{\alpha})^{n}\sim q^{+\frac{1}{2}}$, $n=0,1,2$, while those in red are proportional to $\mathfrak{h}_i\mathfrak{g}_{\alpha}\mathfrak{g}_{\beta}\sim q^{-\frac{3}{2}}$, $\mathfrak{h}_i\mathfrak{h}_j\mathfrak{h}_k\mathfrak{g}_{\alpha}\mathfrak{g}_{\beta}\sim q^{-\frac{1}{2}}$.} 

We obtained \eqref{dS} from the derivative of the defect partition function. This derivative is ``cut-opening" the vacuum diagrams. A technical remark is that, in principle, to $\mathcal{O}(g_{\alpha}g_{\beta})$, this ``cut-opening" of diagrams could be done in two ways, as shown in fig. \ref{fig:cut} below. These correspond to the two possible ways of extracting an overall $\frac{1}{|\vec{p}_T|^2}$ when writing the Fourier-transform in \eqref{I's}. The way a) corresponds to the expression of the integrals $I_2$, with ${\cal I}_2$ given by \eqref{calI2}. The way b) does not represent a contribution to the one-point function.

\begin{figure}[h!]
    \centering
    \includegraphics[scale=.25]{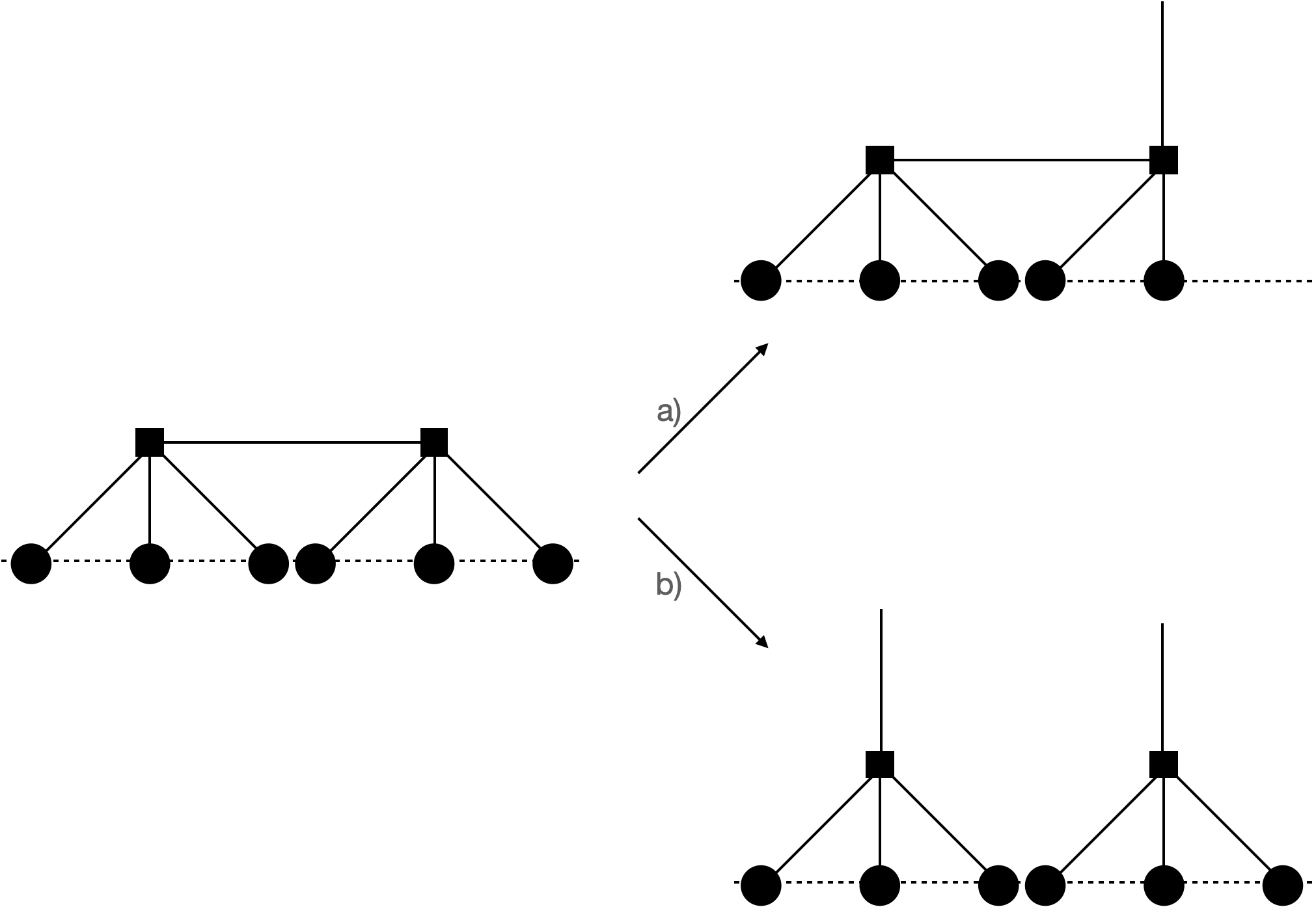}
    \caption{Diagrams contributing to $\partial_i s_{\rm os}$ can be obtained by cutting one line of the vacuum diagram of the defect theory.
    In this example, there are two possible cuts, but only (a) represents a contribution to the one-point function.}
    \label{fig:cut}
\end{figure}

\smallskip

Let us now proceed to the renormalization of the UV divergences by demanding the one-point function to be finite. To get rid of the divergences in \eqref{dS} we introduce the renormalized coupling

\begin{equation}
\nu_{iR}=\nu_i-\frac{ \Omega\,V_i-c^2\,\Omega^2\,V_j\,V_{ij}}{\epsilon}+\frac{\Omega^2\,V_j\,V_{ij}}{2\,\epsilon^2}\,.
\end{equation}
Inverting this equation to quadratic order in the $g_\alpha $'s, one finds 

\begin{equation}
\label{nuai}
\nu_i=\nu_{iR}+\frac{ \Omega\,V^R_i-c^2\,\Omega^2\,V^R_j\,V^R_{ij}}{\epsilon}+\frac{\Omega^2\,V^R_{ij}\,V^R_j}{2\,\epsilon^2}\,,
\end{equation}
where the superscript $R$ stands for evaluation in $\nu_{iR}$.

Having defined the renormalized coupling $\nu_{iR}$, the coefficient of $1/|\vec{p}_T|^2$ is rendered finite. It should be noted that there are still potential divergences in \eqref{dS} and no other parameter to take care of them. However, by writing $\nu_i$ in terms of the renormalized coupling one finds the relation

\begin{equation}
 \Omega\,V_i -\frac{\Omega^2\,V_j\,V_{ij}}{\epsilon}= \Omega\,V^R_i\,,
\end{equation}
Thus, the coefficient of $(\log|\vec{p}_T|)/|\vec{p}_T|^2$ is  made finite by the  same renormalization prescription.

\smallskip

Let us now compute the $\beta$ functions. Restoring the renormalization scale, we have

\begin{equation}
\label{ffi}
\nu_i=\mu^{\frac{\epsilon}{2}}\,(\nu_{iR}+F_i)\,,\qquad F_i\equiv\frac{1}{\epsilon}\big( \Omega\,V^R_i-c^2\,\Omega^2\,V^R_j\,V^R_{ij}\big)+\frac{1}{2\,\epsilon^2}\Omega^2\,V^R_j\, V^R_{ij}\, .
\end{equation}
Using that the bare couplings $\nu_i$ are independent of $\mu$,
we obtain the relation

\begin{equation}
\mu\frac{d\nu_i}{d\mu}=0\quad\implies\quad \frac{\epsilon}{2}\,\nu_{iR}+\frac{\epsilon}{2}\,F_i+\beta_i+\frac{\partial F_i}{\partial \nu_{jR}}\,\beta_j+\frac{\partial F_i}{\partial g_{\alpha} }\,\beta_{g_{\alpha} }=0\,.
\end{equation}

Because  bulk loops are suppressed, the bulk couplings $g_\alpha$ are not renormalized and the corresponding $\beta_{g_{\alpha} }$ 's
are just given by their classical expressions
$\beta_{g_{\alpha} }=-c\,\epsilon\,g_{\alpha} $, determined
by dimensional analysis. 
Using also the fact that the $p$-th derivative $V^R_{i_1\cdots i_p}$ is linear in the $g_\alpha $'s, it follows that

\begin{equation}
\frac{\partial F_i}{\partial g_{\alpha} }\,\beta_{g_{\alpha} }=-c\,\epsilon\,F_i+c\,\Omega^2\,V_j^R\,V_{ij}^R\,\Big( c^2-\frac{1}{2\,\epsilon}\Big)\,.
\end{equation}
Thus

\begin{equation}
\label{betadef}
 \frac{\epsilon}{2}\,\nu_{iR}+\beta_i+\frac{\partial F_i}{\partial \nu_{jR}}\,\beta_j-\epsilon\,\Big(c-\frac{1}{2}\Big)\,F_i+c\,\Omega^2\,V_j^R\,V_{ij}^R\,\Big(c^2-\frac{1}{2\,\epsilon}\Big)=0\,.
\end{equation}
On dimensional grounds, one has

\begin{equation}
\beta_i=-\frac{\epsilon}{2}\nu_{iR}+\beta_i^{(Q)}\,,
\end{equation}
where $\beta_i^{(Q)}$ represents the quantum contributions.
Substituting this expression into \eqref{betadef}, we find the following equation for $\beta_i^{(Q)}$

\begin{equation}
\beta_i^{(Q)}+\frac{\partial F_i}{\partial \nu_{jR}}\,\beta_j^{(Q)}=\epsilon\,\Big(c-\frac{1}{2}\Big)\,F_i+\frac{\epsilon}{2}\nu_{jR}\frac{\partial F_i}{\partial \nu_{jR}}-c\,\Omega^2\,V_j^R\,V_{ij}^R\,\Big(c^2-\frac{1}{2\,\epsilon}
\Big)\,.
\end{equation}
The $p$-th derivative of $V^R$ is a homogeneous polynomial of degree $n-p$ in $\nu_{iR}$. Therefore 

\begin{equation}
\label{homopoli}
\nu_{jR}\frac{\partial V^R_{i_1\cdots i_p}}{\partial\nu_{jR}}=(n-p)V^R_{i_1\cdots i_p}\, ,
\end{equation}
and

\begin{equation}
\nu_{jR}\frac{\partial F_i}{\nu_{jR}}=(n-1)\,F_i-(n-2)\,\Omega^2\,V^R_j\,V^R_{ij}\,\Big(\frac{c^2}{\epsilon}-\frac{1}{2\,\epsilon^2}\Big)\, .
\end{equation}
Thus

\begin{equation}
\beta_i^{(Q)}+\frac{\partial F_i}{\partial \nu_{jR}}\,\beta_j^{(Q)}= 2\,c\,\Omega\,V^R_i-4c^3\,\Omega^2\,V^R_j\,V^R_{ij}+\frac{2\,c}{\epsilon}\,\Omega^2 V^R_j\,V^R_{ij}\,.
\end{equation}
Using the definition of $F_i$ in \eqref{ffi}, 
we get 
\begin{equation}
\frac{\partial F_i}{\partial \nu_{jR}}\,\beta_j^{(Q)}=\frac{ \Omega\,V_{ij}^R}{\epsilon}\,\beta_j^{(Q)}+...\,,
\end{equation}
where we neglected $O(g^3)$ terms (recall that $\beta_j^{(Q)}$ is of order $g$, and every factor $V_{i...}^R$ adds a power of $g$). Therefore 

\begin{equation}
\beta_i^{(Q)}+\frac{ \Omega\,V_{ij}^R}{\epsilon}\,\beta_j^{(Q)}= 2\,c\,\Omega\,V^R_i-4c^3\,\Omega^2\,V^R_j\,V^R_{ij}+\frac{2\,c}{\epsilon}\,\Omega^2\,V^R_j\,V^R_{ij}\,.
\end{equation}
The solution to this equation is

\begin{equation}
\beta_i^{(Q)}=2\,c\,  \Omega\,V^R_i-4c^3\,\Omega^2\,V^R_j\,V^R_{ij}\,.
\end{equation}
Thus, restoring the classical piece, the $\beta $ functions are

\begin{equation}
\label{betaii}
\beta_i=-\frac{\epsilon}{2}\nu_{iR}+2\,c\,   \Omega\,V^R_i-4c^3\,\Omega^2\,V^R_j\,V^R_{ij}\,.
\end{equation}
We note that the $\beta $ functions  can be written as
($\vec{\partial}=\frac{\vec{\partial}}{\partial \vec{\nu}_R}$)

\begin{equation}
\label{betas}
\vec{\beta}=2\,c\,\vec{\partial}\mathcal{H}\,,\qquad \mathcal{H}=-\frac{\epsilon}{8\,c}\nu_{iR}^2+  \Omega\,V^R-c^2\,\Omega^2\,(V^R_{i})^2\,.
\end{equation}
As a check, the $\beta$ functions in \eqref{betas} reproduce the results in \cite{Allais:2014fqa,Cuomo:2021kfm,Rodriguez-Gomez:2022gbz} when particularized to the corresponding models.

Eq.\eqref{betas} shows that the RG flow is a gradient flow of the ''energy function" $\mathcal{H}$ in both in $d=4-\epsilon $ and in $d=6-\epsilon$. Interestingly, the effect of the classical contribution to the $\beta$ functions is to generate a negative mass squared contribution to the ``energy" $\mathcal{H}$.  If the potential $V^R$ (and consequently $(V_i^R)^2$) has a global symmetry, then the theory at the fixed points will exhibit a symmetry breaking pattern similar to that of the Higgs mechanism. For instance, if the theory has an $O(N)$ symmetry, then, at non-trivial fixed points  only an $O(N-1)$ subgroup will be unbroken.\footnote{This property was implicitly used in \textit{e.g.} \cite{Allais:2014fqa,Cuomo:2021kfm,Rodriguez-Gomez:2022gbz}.} 

\subsection{Relation with the circular defect in $d=4-\epsilon $}

The previous computation can be adapted to circular defects of radius $R$ in $d=4-\epsilon$. The on-shell action is 

\begin{equation}
\label{scirc}
S_{\rm os}^{\circ}=-\frac{1}{2}\,\nu_i^2\,I_0+ V\,I_1-\frac{V_i^2}{2}\,I_2\,,
\end{equation}
where now the integrals are to be computed with a $\delta_T$ function localizing not to a line defect but  to the circular defect. To one-loop order, these integrals can be borrowed from \cite{Cuomo:2021kfm,Popov:2022nfq}. Explicitly,

\begin{equation}
I_0=-\frac{\epsilon}{4}\,,\qquad I_1=-2\,\Omega\,.
\end{equation}
Writing \eqref{scirc} in terms of the renormalized coupling and using \eqref{homopoli}, for $n=4$, $c=1$ we find the remarkable formula

\begin{equation}
\label{SSHH}
S_{\rm os}^{\circ}=-\mathcal{H}\,.
\end{equation}
Thus $\exp(\mathcal{H})$ represents the VEV of the 
circular defect.
The formula  $\vec{\beta}=2\vec{\partial}\mathcal{H}=-2\vec{\partial} S_{\rm os}^{\circ} $ agrees with  general arguments given  in \cite{Cuomo:2021rkm,Cuomo:2021kfm,Beccaria:2017rbe}, which applies to  line defects. For the surface defect in 6d,  we found the gradient formula \eqref{betas}, but the analogous interpretation of $\mathcal{H}$ as the on-shell action of a compact defect remains to be understood (for additional  comments, see section 6).

\section{One-point function at fixed points}\label{profile}

Having renormalized the UV divergences, we can now write \eqref{dS} in terms of the renormalized couplings:

\begin{eqnarray}
\label{dSRi}
\partial_is_{\rm os}&=&-\nu_{iR}\,\Big[\int \frac{d\vec{p}_T}{(2\pi)^{d_T}}\frac{1}{|\vec{p}_T|^2}\Big(1+\frac{(C-\Omega\,A)\,V_j^R\,V_{ij}^R}{\nu_{iR}}-\frac{ V_i^R\,\epsilon\,A}{\nu_{iR}}\Big) \\ \nonumber && +\,\int \frac{d\vec{p}_T}{(2\pi)^{d_T}}\,\frac{1}{|\vec{p}_T|^2}\,\Big(\frac{2\,c\, \Omega\,V^R_i -4\,c^3\,\Omega^2\,V_j^R\,V_{ij}^R}{\nu_{iR}}\Big)\,\log|\vec{p}_T| \\ \nonumber &&+\int \frac{d\vec{p}_T}{(2\pi)^{d_T}}\,\frac{1}{|\vec{p}_T|^2}\,\Big( \frac{2\,c^2\,\Omega^2\,V^R_j\,V^R_{ij}}{\nu_{iR}}-\frac{ 2\,c^2\,\Omega\,V_i^R\,\epsilon}{\nu_{iR}}\Big)(\log|\vec{p}_T|)^2\Big]\,.
\end{eqnarray}
Here we see the relevance of keeping the $O(\epsilon)$ term in the integral \eqref{iuno}: it contributes to the finite coefficient
in the $V^R_j\,V^R_{ij}$ term multiplying $(\log|\vec{p}_T|)^2$.
The $\epsilon$ factor canceled out against a $1/\epsilon$ factor coming from a term in the formula \eqref{nuai} for $\nu_i$.

We  shall also keep  terms  proportional to $\epsilon $ in  \eqref{dSRi}, as they will be important when evaluating this expression at the defect fixed point, that is, at the solution of 

\begin{equation}
\label{fijonu}
\beta_i=-\frac{\epsilon}{2}\nu_{iR}+2\,c\   \Omega\,V^R_i-4c^3\,\Omega^2\,V^R_j\,V^R_{ij}=0\,.
\end{equation}
There is a fixed point at $\nu_{iR}=0$ and non-trivial fixed points where
 $g\,(\nu^R)^{n-2}$ are of order $\epsilon$. 
 Here the order $\epsilon$ terms in  \eqref{dSRi} are also important, since, at the fixed point, they will contribute just like the $V^R_j\,V^R_{ij}$ terms, giving rise to $O(\epsilon^2) $ contributions.\footnote{It is easy to check that terms with higher powers of $\epsilon $ in the integrals \eqref{iuno} and \eqref{idose}
 do not  contribute to the order $O(\epsilon^2) $ studied here.}

Let us now compute \eqref{dSRi} at a non-trivial fixed point. The fact that $g\,(\nu^R)^{n-2}\sim \epsilon$ implies that the factor  
multiplying $\frac{1}{|\vec{p}_T|^2}$ in the first line of \eqref{dSRi} 
 is of the form $1+\epsilon^2$. Hence, to order $\epsilon^2$ we can write

\begin{eqnarray}
\partial_is_{\rm os}&=&-\nu_{iR}\,\Big(1+\frac{(C-\Omega\,A)\,V_j^R\,V_{ij}^R}{\nu_{iR}}-\frac{ V_i^R\,\epsilon\,A}{\nu_{iR}}\Big) \\ \nonumber && \int \frac{d\vec{p}_T}{(2\pi)^{d_T}}\,\frac{1}{|\vec{p}_T|^2}\,\Big[ 1 + \Big(\frac{2\,c\, \Omega\,V^R_i -4\,c^3\,\Omega^2\,V_j^R\,V_{ij}^R}{\nu_{iR}}\Big)\,\log|\vec{p}_T| \\ \nonumber &&+\Big( \frac{2\,c^2\,\Omega^2\,V^R_j\,V^R_{ij}}{\nu_{iR}}-\frac{ 2\,c^2\,\Omega\,V_i^R\,\epsilon}{\nu_{iR}}\Big)(\log|\vec{p}_T|)^2\Big]\,.
\end{eqnarray}
Squaring the $\beta $ function equation \eqref{fijonu} and taking its derivative, to quadratic order in the $g_\alpha $'s (or, equivalently, to order $\epsilon^2$), one finds the relation

\begin{equation}
2\beta_j\,\partial_i\beta_j+n\,\epsilon\,\beta_i=(1-n)\,\frac{\epsilon^2}{2}\nu_{iR}+8\,c^2\,(1+c^2\,\epsilon)\,\Omega^2\,V^R_j\,V^R_{ij}\,.
\end{equation}
Thus, at the fixed point, and to order $\epsilon^2$, we finally obtain

\begin{equation}
\label{granfinale}
\partial_is_{\rm os}=-\nu_{iR}\,\mathcal{N}  \int \frac{d\vec{p}_T}{(2\pi)^{d_T}}\,\frac{1}{|\vec{p}_T|^2}\,\Big[ 1 +\frac{\epsilon}{2}\,\log|\vec{p}_T|+\frac{1}{8}(1-2c)\,\epsilon^2\,(\log|\vec{p}_T|)^2\Big]\,, 
\end{equation}
with 

\begin{equation}
    \mathcal{N}=\Big(1+\frac{(C-\Omega\,A)\,V_j^R\,V_{ij}^R}{\nu_{iR}}-\frac{ V_i^R\,\epsilon\,A}{\nu_{iR}}\Big)=1+\frac{(1+2c)\,C-(1+6c)\,\Omega\,A}{16\,c^2\,\Omega^2}\,\epsilon^2\,.
\end{equation}
In the last equality we used that we are sitting at a non-trivial fixed point. The constant $\mathcal{N}$ is scheme dependent.

The one-point function \eqref{granfinale} shows a deviation from the conformal behavior at order $\mathcal{O}(\epsilon^2)$. 
One formally recovers the CFT behavior by setting $c\to 0$.
This is consistent with the fact that $c$ appears as a coefficient in the bulk coupling $\beta$ function. Recall that in the double-scaling limit studied here  the running of bulk couplings is completely determined by the classical contribution to the $\beta $ functions, dictated by dimensional analysis.

\subsection{General predictions from the Renormalization Group}

It is possible to understand the formula \eqref{granfinale} from a deeper perspective. Let us consider the quantity $\hat{G}_i\equiv \partial_is_{\rm os}$. Explicitly
 
\begin{equation}
\label{Ghat1}
    \hat{G}_i=-\mathcal{V}^{-1}\,\int d^dx\,\langle \phi\rangle \,\delta_T(\vec{x})\,.
\end{equation}
Since $\hat{G}_i$ has dimension $d_{\rm wv}-\frac{\epsilon}{2}$ and must vanish when $\nu_{iR}=0$, it can be written as  

\begin{equation}
    \hat{G}_i=-\nu_{iR}\, \int \frac{d\vec{p}_T}{(2\pi)^{d_T}}\,\frac{1}{|\vec{p}_T|^{2}}\,\hat{F}(\frac{|\vec{p}_T|}{\mu},\mu^{n-c\, d}\,g_{\alpha},\,\mu^{-\frac{\epsilon}{2}}\,\nu_{iR})\,,
\end{equation}
where $\hat{F}$ is a dimensionless function of its arguments.  

Scalar fields do not get  anomalous dimension because loops of the bulk coupling are suppressed in the double-scaling limit.
Then, the function $\hat{G}_i$ satisfies the Callan-Symanzik equation

\begin{equation}
\label{callansym}
    \mu\,\frac{d\hat{G}_i}{d\mu}=-\frac{\epsilon}{2}\,\hat{G}_i\,.
\end{equation}
Since $\hat{G}_i$ is a function of $\{\mu,\,g_{\alpha}(\mu),\,\nu_{iR}(\mu)\}$, this equation becomes

\begin{equation}
\label{callsymhatG}
    \mu\frac{\partial \hat{G}_i}{\partial\mu}+\frac{\epsilon}{2}\,\hat{G}_i+\beta_{g_{\alpha}}\,\frac{\partial \hat{G}_i}{\partial g_{\alpha}}+\beta_j\,\frac{\partial \hat{G}_i}{\partial \nu_{jR}}=0\,.
\end{equation}
We can cross-check these equations, in particular, the RHS in \eqref{callansym}, in the case of a dCFT, where $\beta_{g_{\alpha}}=\beta_i=0$.  Then, the Callan-Symanzik equation \eqref{callsymhatG} gives $\hat{F}=(|\vec{p}_T|/\mu)^{\frac{\epsilon}{2}}$, leading to

\begin{equation}
    \hat{G}_i=-\nu_{iR}\,\mu^{-\frac{\epsilon}{2}}\, \int \frac{d\vec{p}_T}{(2\pi)^{d_T}}\,\frac{1}{|\vec{p}_T|^{2-\frac{\epsilon}{2}}}\,.
\end{equation}
This can be written as

\begin{equation}
    \hat{G}_i=-\mathcal{V}^{-1}\, \int dx\, \langle\phi\rangle\,\delta_T(\vec{x})\,,\qquad \langle \phi\rangle = \nu_{iR}\,\mu^{-\frac{\epsilon}{2}}\int \frac{d\vec{p}_T}{(2\pi)^{d_T}}\,\frac{e^{i\vec{p}_T\cdot\vec{x}_T}}{|\vec{p}_T|^{2-\frac{\epsilon}{2}}}
\end{equation}
which is the expected form for the 1-point function (compare with \textit{e.g.} (21) in \cite{Allais:2014fqa} in the case of vanishing anomalous dimension).

Coming back to our case, in perturbation theory

\begin{equation}
    \hat{F}=\sum_{k=0}^{\infty}\,(\mu^{n-c\, d}\,g_{\alpha} )^k\,\hat{f}_k(\frac{|\vec{p}_T|}{\mu},\,\mu^{-\frac{\epsilon}{2}}\,\nu_{iR})\,.
\end{equation}
Here $(g_\alpha)^k$ is a symbolic notation for terms involving $k$ of the $g$'s, such as $g_{\alpha_1}...g_{\alpha_k}$.\footnote{ A more proper notation would be to write $\hat{F} = \sum_{k=0}^{\infty}\sum_{\alpha_1,\cdots \alpha_k}(\mu^{n-c\, d}\,g_{\alpha_1}) \cdots (\mu^{n-c\, d}\,g_{\alpha_k}) \hat{f}_k^{\alpha_1\cdots\alpha_k}$. We choose to condense the notation in the obvious way in order not to clutter the presentation.} Then

\begin{equation}
    g_{\alpha} \frac{\partial \hat{f}}{\partial g_{\alpha} }=\sum_{k=1}^\infty\,k\,\hat{F}_k\,,\ \qquad \hat{F}_k\equiv (\mu^{n-c\, d}\,g_{\alpha} )^k\,\hat{f}_k\ .
\end{equation}
 Using that $\beta_{g_{\alpha} }=-c\,\epsilon\,g_{\alpha} $, we get

\begin{equation}
   \beta_{g_{\alpha} }\frac{\partial  \hat G_i}{\partial g_{\alpha} }=-\nu_{iR}\, \int \frac{d\vec{p}_T}{(2\pi)^{d_T}}\,\frac{1}{|\vec{p}_T|^{2}}
   \,\sum_{k=1}^\infty\,(-c\,\epsilon)\,k\,\hat{F}_k\, .
\end{equation}
Thus, at the defect fixed point, the Callan-Symanzik equation \eqref{callsymhatG} becomes

\begin{equation}
\label{afsa}
    \sum_{k=0}^\infty \Big(\mu \frac{\partial \hat{F}_k}{\partial\mu} +\frac{\epsilon}{2}\hat{F}_k \Big)-c\, \epsilon\,\sum_{k=1}^\infty\,k\,\hat{F}_k=0\,.
\end{equation}
At the defect fixed point, the $\nu_{iR}$ are solutions of $\beta_i=0$, given in \eqref{fijonu}. In the $\epsilon$ expansion, the solution is of the form
$g_\alpha \nu_i\nu_k =O(\epsilon)$ in the 4d theory (see also section 6.1 below) and $g_\alpha \nu_i =O(\epsilon)$ in the 6d theory.
Since $\hat{F}_k\propto (g_\alpha)^k$, it follows that, at the defect fixed point, $\hat F_k\sim \epsilon^k$. More generally, it is of the form $\hat{F}_k=\epsilon^k\,\sum_{r=0} a_{k,r}\,\epsilon^r$\,. Thus, we see that the last term in \eqref{afsa}  contributes quadratically in $\epsilon$. Consequently, the first two terms of $\hat{F}$ coincide with those of a conformal theory. 

Explicitly, to order $\epsilon^0$, $\epsilon^1$ and $\epsilon^2$, \eqref{afsa} yields to

\begin{equation}
   \mu \frac{\partial \hat{F}_0}{\partial\mu}=0\,, \qquad \mu \frac{\partial \hat{F}_1}{\partial\mu}  +\frac{\epsilon}{2}\hat{F}_0=0\,,\qquad
       \mu \frac{\partial \hat{F}_2}{\partial\mu}+\frac{\epsilon}{2}\hat{F}_1 -c\, \epsilon\,\hat{F}_1=0\,.
\end{equation}

Appropriately redefining the integration constants the solution is

\begin{equation}
    \hat{F}=B+\frac{\epsilon}{2}\,\log\big(\frac{|\vec{p}_T|}{\mu}\big)+\frac{1-2c}{8}\,\epsilon^2 \Big(\log\big(\frac{|\vec{p}_T|}{\mu}\big)\Big)^2+\frac{b_1}{2}(1-2c)\epsilon^2 \log\big(\frac{|\vec{p}_T|}{\mu}\big)+ \mathcal{O}(\epsilon^3)\,,
\end{equation}
where $B=1+b_1\epsilon+b_2\epsilon^2 $ and $b_1$, $b_2$ are integration constants.
It is convenient to factorize  $B$, which then
appears in the overall normalization.
Thus, to this order

\begin{equation}
    \hat{F}=1+\frac{\epsilon}{2}\,\log\big(\frac{|\vec{p}_T|}{\mu}\big)+\frac{1-2c}{8}\,\epsilon^2 \Big(\log\big(\frac{|\vec{p}_T|}{\mu}\big)\Big)^2-b_1\,c\,\epsilon^2 \log\big(\frac{|\vec{p}_T|}{\mu}\big)+ \mathcal{O}(\epsilon^3)\,.
\end{equation}
This  reproduces our
previous result \eqref{granfinale},
obtained by explicit calculation in perturbation theory (with the integration constant $b_1=0$). In order to compute higher order terms in the $\epsilon$ expansion, one would need to compute $\beta_i$ to order $g_\alpha^3$.

As stressed earlier,  the first two terms of $\hat{F}$ coincide with those of a conformal theory. For this observable, the deviation from conformality is seen at the order $O(\epsilon^2)$.

\section{RG flows in the Twins model}\label{twins}

\subsection{Fixed points and RG flows}

As an application, here we study a $\mathbb{Z}_2\times \mathbb{Z}_2$-symmetric model with two scalar fields
 in $d=4-\epsilon $, defined by the action (hereafter, the ``twins" model):

\begin{equation}
\mathcal{S}= q\int d^d x\left( \frac{1}{2}\big(\partial\phi_1\big)^2+\frac{1}{2}\big(\partial\phi_2\big)^2+V(\phi_1,\phi_2)\right)\, , 
\end{equation}
with
\begin{equation}
    V(\phi_1,\phi_2)=\frac14 g_1\phi_1^4+\frac14 g_2\phi_2^4+\frac12 g_3 \phi_1^2\phi_2^2\ .
\end{equation}
The model has a $\mathbb{Z}_2\times \mathbb{Z}_2$ symmetry under independent transformations $\phi_1\to -\phi_1$, and
$\phi_2\to -\phi_2$. When $g_1=g_2$ the theory is also symmetric under the exchange $\phi_1\leftrightarrow \phi_2$. At the special point $g_1=g_2=g_3$, the theory has $O(2)$ symmetry.

 We assume that the potential is bounded from below, which requires  $g_1\geq 0$, $g_2\geq 0$ and $g_3\geq -\sqrt{g_1 g_2}$.
 In the present case, this also implies that the potential is positive semi-definite.
When $g_3=\pm \sqrt{g_1 g_2}$, the potential is a perfect square. If $g_3=-\sqrt{g_1 g_2}$, then the potential has flat directions $\phi_2=\pm \phi_1 (g_1/g_2)^\frac14 $.

\medskip

 We now consider the action deformed by the defect,  

\begin{equation}
\mathcal{S}= q\int d^d x\left( \frac{1}{2}\big(\partial\phi_i\big)^2+V(\phi_i)+\nu_i\,\phi_i\,\delta_T(\vec{x})\right)\,, \quad i=1,2\ .
\end{equation}
Just as in the general model \eqref{esses} with $N$ scalar fields, in the $q\to\infty$ limit  all bulk loop diagrams of the twins model are suppressed, leaving
only the  diagrams computed in the previous sections.
We can directly apply the previous results to write down the $\beta $ functions and the VEV's of $\phi_1$ and $\phi_2$.

Setting $n=4$, $c=1$, from  \eqref{betaii} we have
\begin{eqnarray}
\beta_{\nu_1}&=& \nu_{1R}\Big( -\frac{\epsilon}{2}+2\, \,\Omega\,(g_1(\nu_{1R})^2+g_3(\nu_{2R})^2)
\nonumber\\
&-&
4\,\Omega^2\, \big[ 3g_1^2(\nu_{1R})^4 +2g_3(2g_1+g_3)(\nu_{1R})^2(\nu_{2R})^2+g_3(2g_2+g_3) (\nu_{2R})^4\big]\Big)\, ,
\end{eqnarray}
\begin{eqnarray}
\beta_{\nu_2}&=& \nu_{2R}\Big( -\frac{\epsilon}{2}+2\, \,\Omega\,(g_2(\nu_{2R})^2+g_3(\nu_{1R})^2)
\nonumber\\
&-&
4\,\Omega^2\, \big[ 3g_2^2(\nu_{2R})^4 +2g_3(2g_2+g_3)(\nu_{1R})^2(\nu_{2R})^2+g_3(2g_1+g_3) (\nu_{1R})^4\big]\Big)\,.
\end{eqnarray}
Let us first identify the fixed points. It is convenient to define
\begin{equation}
 x_1=\Omega\, g_1 \,\nu_{1R}^2\ ,\quad   x_2=\Omega\, g_2 \,\nu_{2R}^2\ ,\qquad \zeta=\frac{g_3}{g_1}\ ,\ \ 
  \eta=\frac{g_3}{g_2}\ .
\end{equation}
Note that the parameters $\zeta$ and $\eta$
do not run, since $g_1, g_2, g_3$ have the same classical RG flow,
{\it i.e.} $\beta_{g_\alpha}=-\epsilon g_\alpha$.

Fixed points can be divided into two classes. First of all, we have three fixed points which do not depend on the parameters $\zeta,\, \eta$. They are located at $(x_1^*,x_2^*)$ given by

\begin{equation}
\label{fijoabb}
 (x_1^*,x_2^*):\ \qquad    a)\ \,(0,\,0), \qquad b)\ \,(0,\,\frac{1}{4}\,(\epsilon+\frac{3}{2}\epsilon^2)) \qquad b')\ \,(\frac{1}{4}\,(\epsilon+\frac{3}{2}\epsilon^2),\,0)\,.
\end{equation}
In addition to these fixed points, there is  a fixed point that  depend on the parameters, located at

\begin{equation}
\label{fijoc}
    c)\ \, \left(\frac{1-\eta}{4(1-\zeta\eta)}\,(\epsilon+\frac32 \epsilon^2),\, \frac{1-\zeta}{4(1-\zeta\eta)}\,(\epsilon+\frac32 \epsilon^2) \right)\ .
\end{equation}
It should be noted that existence of the fixed points requires $x_1^*\geq 0$ and $x_2^*\geq 0$ to ensure that the $\nu_{iR}$'s are real. Here we assume
 $\epsilon>0$ (if $\epsilon <0$ there is only one fixed point at the origin $(x_1^*,x_2^*)=(0,0)$ and is attractive). Note as well that in the original  $(\nu_{1R},\nu_{2R})$ variables, fixed points $b,\  b'$ and $c$ give rise to a pair of ``mirror" fixed points, defined by the two sign choices  $\nu_{iR}\propto \pm \sqrt{x_i}$.

Before further proceeding, let us pause to stress that in the double-scaling limit we are taking the couplings in the original lagrangian are scaled as  $\mathfrak{g}_\alpha\to 0$ as $q^{-1}$. As a result, bulk loop diagrams are suppressed (as expected for a semiclassical limit, since $q^{-1}$ plays the role of $\hbar$). This is in particular reflected in the $\beta$ functions for the bulk couplings, which are of the generic form $\beta_{\mathfrak{g}_{\alpha}}\sim -\epsilon\,\mathfrak{g}_{\alpha}+O(\mathfrak{g}_{\alpha}^3)$. When rewritten in terms of the re-scaled couplings, these become  $\beta_{g_{\alpha}}\sim -\epsilon\,g_{\alpha}+O(\frac{g_{\alpha}^3}{q^2})\to \,-\epsilon\,g_{\alpha}$. Therefore, the  $\beta_{g_{\alpha}} $ functions are simply the classical ones: in this limit, the bulk couplings run solely because of classical dimensional analysis (in particular, note that in this limit one cannot sit in the WF fixed point).\footnote{
An interesting
problem is  to incorporate further quantum corrections in order to study the theory at both bulk and defect fixed points.  Demanding that $\beta_{g_{\alpha}}$ functions vanish, one has  $g_\alpha^2\sim \epsilon q^2$. 
Thus, at large $q$,  bulk fixed points would lie at strong coupling. A systematic organization of perturbation theory may be attempted by choosing $\epsilon\sim 1/q^2$. We leave this  problem for future studies.} Thus, 
while the fixed points occur at small values $g\nu^2 =\mathfrak{g}\mathfrak{h}^2=O(\epsilon)$,  bulk loop corrections  come with extra powers of $q^{-1}$, and thus are negligible with respect to the diagrams that we have considered.\footnote{
The double-scaling limit selects the same diagrams as the ``ladder" approximation in \cite{Beccaria:2017rbe,Beccaria:2021rmj}.}

IR stability can be studied by linearization about a given fixed point. The defect RG flow is a gradient (in the $\nu_{iR}$'s) flow of the ``energy function" $-\mathcal{H}$ when one uses the RG time variable $t=-\log\mu$. Computing the eigenvalues $(\lambda_1,\lambda_2)$ of the Hessian at each fixed point one finds (here we work to order $\epsilon$)

\begin{equation}
 (\lambda_1,\lambda_2):\qquad   a)\, \ (\frac{\epsilon}{2},\,\frac{\epsilon}{2}),\qquad b)\,\ (-\epsilon,\,\frac{\epsilon}{2}(1-\eta)),\qquad b')\,\  (-\epsilon,\,\frac{\epsilon}{2}(1-\zeta))\,,
\end{equation}
and

\begin{equation}
    c)\, \ (-\epsilon\,\frac{(1-\eta)\,(1-\zeta)}{1-\eta\,\zeta},\,-\epsilon)\,.
\end{equation}
A given fixed point is IR stable in all directions if both eigenvalues are negative and it develops an unstable direction when an eigenvalue becomes positive.
Note in particular that the origin, {\it i.e.} fixed point $a$, is always repulsive since we are  assuming $\epsilon>0$.

\medskip

There are  three different regimes: 
\begin{enumerate}

 \item[(i)]   $\zeta,\,\eta<1$. Then $c$ is attractive, whereas $b$ and $b'$ are repulsive.

 \item[(ii)] $\zeta<1$, $\eta>1$. 
Here $b$ is an attractor and $b'$ is repulsive.
There is no fixed point $c$ in this regime, since this exists provided $(1-\eta)\,(1-\zeta)>0$.
[The  case  $\zeta>1$, $\eta<1$ is equivalent under the exchange of $b$ and $b'$.] 

 \item[(iii)] $\zeta,\,\eta>1$. $b$ and $b'$ are  attractors and $c$ is repulsive.

\end{enumerate}

\smallskip

\begin{figure}[h!]
    \centering
    \begin{subfigure}[t]{0.31\textwidth}
        \centering
        \includegraphics[width=\linewidth]{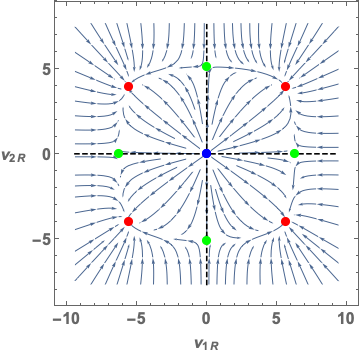} 
        \caption{A case with $\eta<1$, $\zeta<1$. Here $(\zeta,\eta)=(\frac{1}{2},\frac{1}{3})$  .}
 \label{fig:a}
    \end{subfigure}
    \hfill
    \begin{subfigure}[t]{0.31\textwidth}
        \centering
        \includegraphics[width=\linewidth]{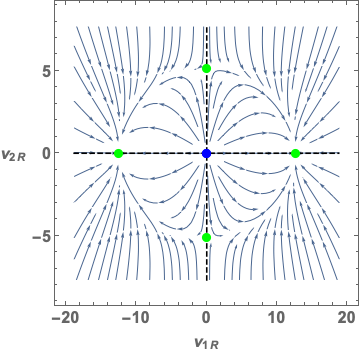} 
        \caption{ $(\zeta, \eta)=(2,\frac{1}{3})$, describing a case with $\eta<1$, $\zeta>1$.} 
        \label{fig:b}
    \end{subfigure}
    \hfill
    \begin{subfigure}[t]{0.31\textwidth}
        \centering
        \includegraphics[width=\linewidth]{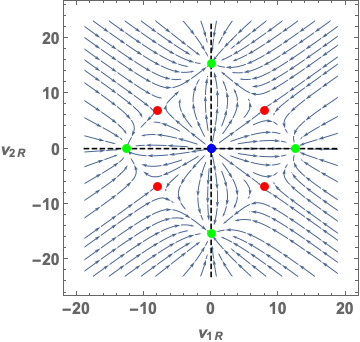} 
        \caption{$(\zeta, \eta)=(2,3)$, describing a case with $\eta>1$, $\zeta>1$.} 
        \label{fig:c}
    \end{subfigure}
    \caption{RG trajectories for representatives of each relevant case.  Blue, green and red dots represent, respectively, the $a$, $\{b,\,b'\}$ and $c$ fixed points. 
    } \label{fig:flows}
\end{figure}

\noindent In fig. \ref{fig:flows} we show the RG flows in each of these regimes. 
Note that regime (i) also includes the cases $\zeta<0$  and $\eta<0 $, corresponding to $g_3<0$.

\begin{figure}[h!]
    \centering
    \includegraphics[scale=.41]{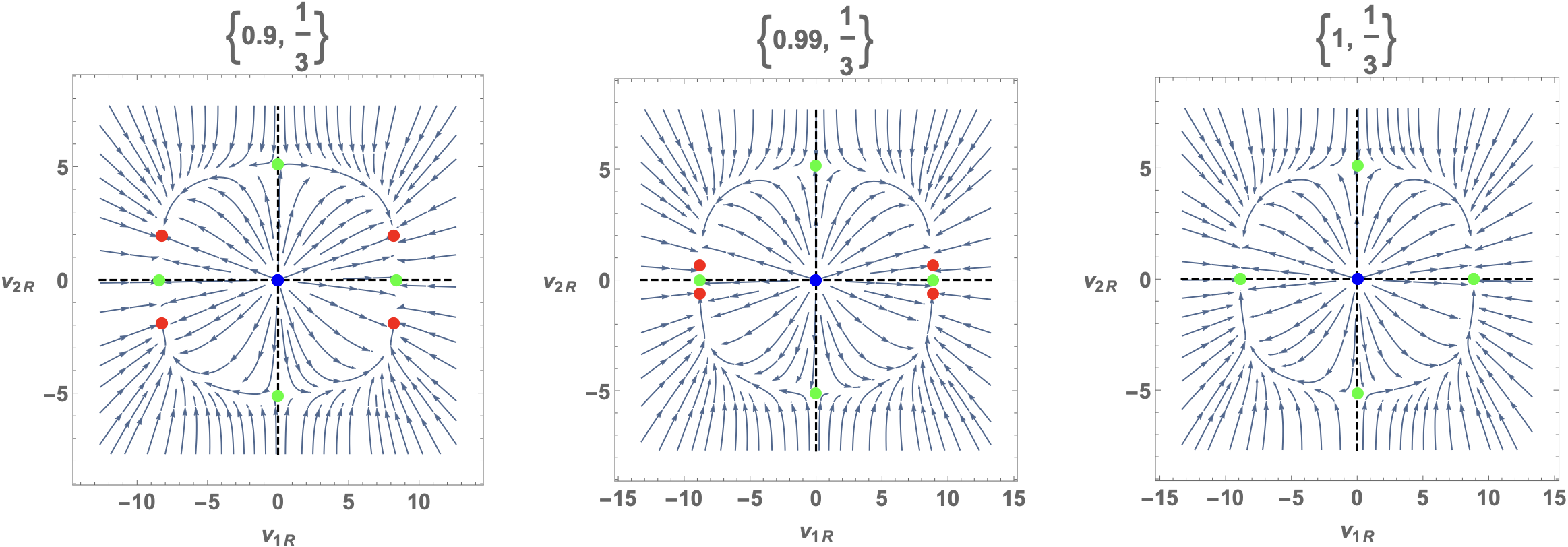}
    \caption{Fixed point annihilation as $\zeta$ grows for fixed $\eta=1/3$. When $\zeta >1$, the RG flow is as illustrated in
    fig. 3 (ii).}
    \label{fig:annihilation}
\end{figure}

Interestingly, as one goes from one regime to another, the fixed point $c$ may disappear or reappear. This happens through fixed point annihilation/creation between two $c$ ``mirror" fixed points.
The annihilation/creation  occurs either at $\nu_{1R}=0$ or $\nu_{2R}=0$, that is, when either $\eta=1$ or $\zeta=1$. At these critical values of the parameters, the
fixed point $c$ merges either with $b$ or with $b'$.
In fig. \ref{fig:annihilation} we show an example of this phenomenon when going from (i) to (ii).

\medskip

The twins model interpolates between the $O(2)$ model
 and the two decoupled $\phi^4$  theories that arise when  $g_3\to 0$.
 The latter has $\eta=\zeta=0$ and the familiar WF fixed points, which
 agree with \eqref{fijoabb} and \eqref{fijoc}, taking into account that there are two possible solutions for each $x_i^*$, the trivial fixed point at the origin and the WF fixed point. 
 
 On the other hand, in the $O(2)$ limit, $\eta\to 1$, $\zeta\to 1$, the fixed point $c)$ is singular and its limiting value
 is ambiguous as it depends on the path the point $\eta=\zeta=1$ is approached.
The general solution to $\beta_1=\beta_2=0$ is given by a circle of fixed points at $\Omega g_1(\nu_{1R}^2+\nu_{2R}^2)=\frac14 \epsilon+\frac38\epsilon^2+\cdots$.

\subsection{Fixed points to all  orders: A Conjecture}

A striking property of the fixed point $c)$,
computed to $O(g^2)$ in \eqref{fijoc},
is that the $\epsilon$ dependence factorizes from the coupling dependence. We will call this phenomenon ``dimensional disentangling".
A natural question is whether this property extends to higher orders or it is just an accident of the first two orders.
Factorization to all orders may seem like a miracle, since at high orders there are a large number of Feynman diagrams  involving  $g_1, g_2$ and $g_3$ in a non-trivial way.
Nevertheless, as we explain below, there are strong indications
that this factorization property must be an exact feature
of the  large $q$ limit, which  is maintained to all order in $\epsilon$. Concretely, we conjecture  that
the fixed point $c$ is exactly located at
\begin{equation}
\label{ccjj}
     c):\quad (x_1^*, x_2^*)=\left(\frac{1-\eta}{4(1-\zeta\eta)}\, f(\epsilon),\, \frac{1-\zeta}{4(1-\zeta\eta)}\, f(\epsilon )\right)\ ,
\end{equation}
and, to all orders, there are other three fixed points $a$, $b$ and $b'$  located at
\begin{equation}
  a)\ \,(0,\,0), \qquad b)\ \,\big(0,\,\frac{1}{4}\,f(\epsilon)\big)\ , \qquad b')\ \, \big( \frac{1}{4}\,f(\epsilon),\,0 \big)\,.
\end{equation}
To justify this conjecture, let us first consider  the fixed points $b$ and $b'$. When either $\nu_{1R}=0$ or $\nu_{2R}=0$, there is only one non-trivial $\beta $ function. This
must be  identical to the quartic model with a single coupling.
Consider, for example, the case $\nu_{2R}=0$.
Because bulk loops are suppressed,  there is no Feynman diagram
involving vertices that depend on $g_2$ or $g_3$, since any $\phi_2$ line could only appear in loops.
Therefore the $\beta $ function must be of the form $\beta_1=\nu_{1R}\,[-\frac{\epsilon}{2}+P(x_1)]$, which ensures that $x_1^*=f(\epsilon)$.

On the other hand, the fixed point $c$ must merge with fixed points $b$ or $b'$ when fixed point annihilation  between mirror fixed points occur, since (due to the $\mathbb{Z}_2\times \mathbb{Z}_2$ symmetry) this always occurs either at $\nu_{1R}=0$ or $\nu_{2R}=0$, and $b$ or $b'$
are the unique solutions with  either $\nu_{1R}=0$ or $\nu_{2R}=0$ (in perturbation theory in $\epsilon$).
This implies  that, if $c$ is of the form  \eqref{ccjj}, exactly the same function $f(\epsilon)$ must appear in the fixed points $b,\ b'$.

In what follows we  present more evidence   for this conjecture:
we will now explicitly show that factorization
holds at order $g^3$, $g^4$ and $g^5$.

Consider computing  computing the 1-point function in terms of Feynman diagrams. The Feynman rules for the vertices are schematically as shown in fig. \ref{fig:FeynmanRules} below.

\begin{figure}[h!]
    \centering
    \includegraphics[scale=.3]{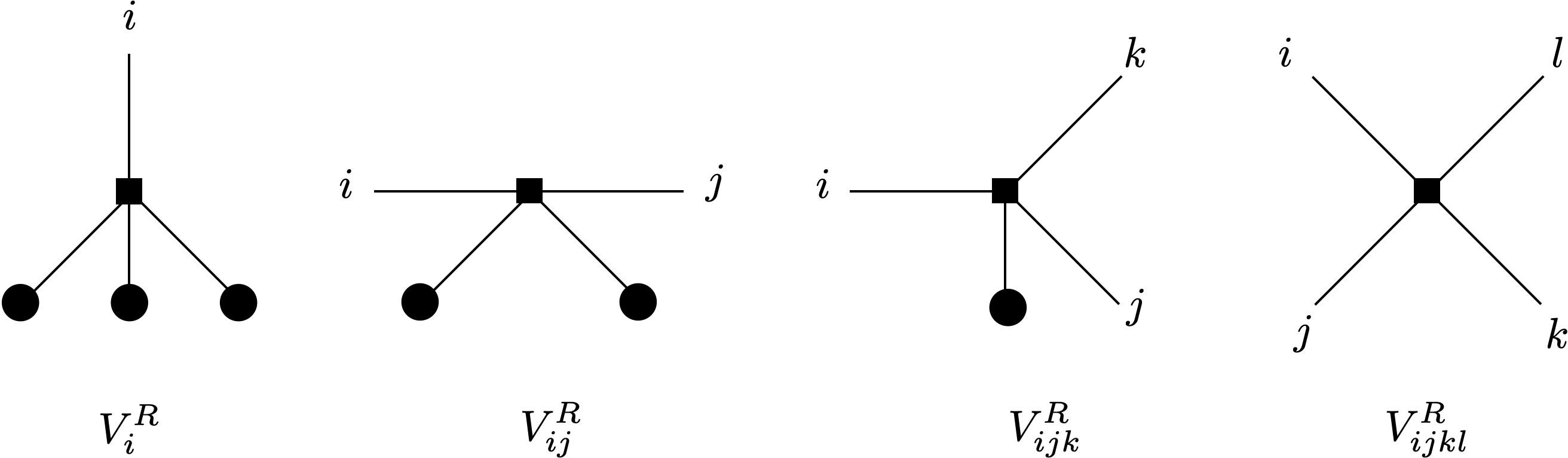}
    \caption{Feynman rules for the vertices (each function is evaluated on the renormalized couplings). Black circles represent the attachment to the defect while squares are bulk vertices. In addition, by virtue of the double-scaling limit,  bulk loops are not allowed and surviving diagrams have 
     the largest number of defect insertions (that is, black circles).}
    \label{fig:FeynmanRules}
\end{figure}
With these rules, one may can construct all possible diagrams to a given order in the couplings (equivalently, in $V^R$). An extra piece of information is that the RG flow must be a gradient flow of the $\mathcal{H}$ function constructed from $V$ and its derivatives evaluated on the renormalized couplings. This makes it possible to further restrict possible contributions to  $\mathcal{H}$, which, to fifth order, can only contain the following terms:

\begin{eqnarray}
    \mathcal{H}&=&-\frac{\epsilon}{4}\,\nu_{iR}^2+2\,\Omega\,V^R-2\,\Omega^2\,(V_i^R)^2\\ \nonumber && +\alpha_1\,\Omega^3\,V_{ij}^R\,V_i^R\,V_j^R\\ \nonumber && +\alpha_2\,\Omega^4\,V_{ijk}^R\,V_i^R\,V_j^R\,V_k^R+\alpha_3\,\Omega^4\,V_{ij}^R\,V_{jk}^R\,V_{k}^R\,V_i^R\\ \nonumber && +\alpha_4\,\Omega^5\,V_{ijkl}^R\,V_i^R\,V_j^R\,V_k^R\,V_l^R+\alpha_5\,\Omega^5\,V_{ijk}^R\,V_{kl}^R\,V_i^R\,V_j^R\,V_l^R+\alpha_6\,\Omega^5\,V_{ij}^R\,V_{jk}^R\,V_{kl}^R\,V_l^R\,V_i^R \\ \nonumber 
    && +\cdots\,,
\end{eqnarray}
where $\alpha_i$ are numerical coefficients which can be determined from higher loop integrals. Using that $\beta_i=2 \partial_i\mathcal{H}$, 
and solving $\beta_1=\beta_2=0$ in perturbation theory, we find that  fixed points $a$, $b$, $b'$ still exist and are located at

\begin{equation}
  a)\ \,(0,\,0), \qquad b)\ \,\big(0,\,\frac{1}{4}\,f(\epsilon)\big)\ , \qquad b')\ \, \big( \frac{1}{4}\,f(\epsilon),\,0 \big)\, ,
\end{equation}
while the $c$ fixed point sits at

\begin{equation}
     c)\ \, \left(\frac{1-\eta}{4(1-\zeta\eta)}\, f(\epsilon),\, \frac{1-\zeta}{4(1-\zeta\eta)}\, f(\epsilon )\right)\ ,
\end{equation}
with

\begin{eqnarray}
    f(\epsilon)&=&\epsilon+\frac{3}{2}\epsilon^2+\frac{3}{4}(6-\alpha_1)\,\epsilon^3+\frac{15}{64}(72-24\alpha_1-2\alpha_2-3\alpha_3)\,\epsilon^4\\ \nonumber && +\frac{9}{128}(1008-504\alpha_1+24\alpha_1^2-60\alpha_2-90\alpha_3-2\alpha_4-6\alpha_5-9\alpha_6)\,\epsilon^5\cdots\,,
\end{eqnarray}
Therefore,  coupling dependence and $\epsilon$ dependence still factorize  to orders $g^3$, $g^4$ and $g^5$.
This strongly indicates that this remarkable property must hold to all orders.

\medskip

Finally, it may be worth noting that the factorization property allows  one to define
$\epsilon'\equiv f(\epsilon)$ so that $d=4-\epsilon' +O\big({\epsilon'}^2\big)$ and the fixed points become of the form
$$
b):\ (0,\frac14\epsilon')\ , \quad b):\ (0,\frac14\epsilon')\ ,
\quad
 c)\ \, \left(\frac{1-\eta}{4(1-\zeta\eta)}\, \epsilon',\, \frac{1-\zeta}{4(1-\zeta\eta)}\, \epsilon'\right)\ .
$$
All higher order corrections get encapsulated in $4-d$ (given by $f^{-1}(\epsilon')$). This redefinition of $\epsilon$ is straightforward if $f(\epsilon )$ is monotonic. A non-monotonic $f(\epsilon )$ could give rise to interesting effects at finite values of $\epsilon$.

\section{Discussion}

Let us summarize the main results of this paper.
We considered a  theory with $N$ interacting scalar fields
with a general (marginal) potential in $d=4-\epsilon$ and $d=6-\epsilon$,
deformed by localized defect (a line in 4d; a surface in 6d).
The deformation is implemented  by an operator which is a linear combination of the scalar fields.
We showed that there is a double-scaling limit where the bulk theory is classical --bulk loops are suppressed-- whereas the defect theory is quantum. This generalizes earlier related studies \cite{Allais:2014fqa,Cuomo:2021kfm,Rodriguez-Gomez:2022gbz}.

The renormalization of the defect theory was here carried out explicitly up to order $g_\alpha^2$, where
the $g_\alpha $'s represent the bulk couplings.  We have  computed the $\beta $ functions of the defect couplings and the VEV's of the
scalar fields on the defect. 
We have  shown that the
$\epsilon$ dependence and general structure of the VEV's at the fixed points are dictated by
the renormalization group, by providing an independent derivation through the Callan-Symanzik equation.

\smallskip

There are a number of  extremely interesting open problems:

\begin{enumerate}
    \item In section 3, we have found  that, to two-loops in the defect couplings, the $\beta$-function of the defect couplings can be written as a gradient, $\partial_i \beta =2c\partial_i \mathcal{H}$, for both the case of a line defect in $d=4-\epsilon$ dimensions and the case of a surface defect in $d=6-\epsilon$ dimensions. It would be very interesting to  prove  more generally that the defect $\beta$ functions  must be  the gradient of a scalar function.
    
    \item For the case of line defects in $d=4-\epsilon$ dimensions, it has been shown in section 3.1 that the function $\exp(\mathcal{H})$ is the VEV of the circular defect, in agreement with \cite{Cuomo:2021rkm,Cuomo:2021kfm,Beccaria:2017rbe}. Similarly, it would be extremely interesting to elucidate the interpretation of the $\mathcal{H}$ function in $d=6-\epsilon$ dimensions as, perhaps, minus the on-shell action of a spherical defect (or other closed surface). 
    
    \item As  shown in section 4, the non-conformality of the bulk theory kicks in at two-loops in the defect couplings in the one-point function. On the face of it, this suggests that the identification of $\exp(\mathcal{H})$ with the VEV of the circular defect  may fail starting from two loops. It would be interesting to investigate this further.
    
    \item The RG flows of the defect couplings depend
    on  RG invariant parameters given by the ratio of bulk couplings.
    This generically leads to an interesting pattern of fixed points which move, annihilate or are created, as the bulk couplings are varied. 
    This was illustrated by considering a line defect in a simple $\mathbb{Z}_2\times \mathbb{Z}_2$ symmetric model with two scalar fields --the ``twins model". The model interpolates between the $O(2)$ theory and two decoupled $\phi^4$ theories, and it can be viewed as the $O(2)$ theory with the addition of a general $O(2)$ breaking deformation preserving $\mathbb{Z}_2\times \mathbb{Z}_2$. A problem of interest is to investigate the effects of fixed point creation/annihilation in other observables, such as {\it e.g.} correlation functions (for which recent results have been obtained \cite{Cuomo:2021kfm, Gimenez-Grau:2022czc}) of defect operators or fusion of defects (such as in \cite{Soderberg:2021kne,Rodriguez-Gomez:2022gbz}). 
    
    \item The fixed points exhibit a remarkable structure where the $\epsilon$-dependence factorizes from the coupling dependence. We argued that dimensional disentangling is a consequence of the double scaling limit and that it must hold to all orders. 
    A corollary of this hypothesis is that the location of fixed points and the properties of the RG flows determined at one loop are the same to all orders in the $\epsilon$ expansion, modulo  an overall change of scale $\epsilon\to f(\epsilon)$.
    An open question is to  understand the underlying reason for this factorization and whether this property  still holds for theories with $N$ scalar fields (a preliminary calculation for three scalar fields suggests that the factorized form of fixed points still holds in this case). It would also be very interesting to see if dimensional disentangling also holds in more general defect theories, in particular, gauge theories with defects, upon taking a suitable  large defect coupling limit.

\end{enumerate}

\section*{Acknowledgements}

D.R-G. thanks the CERN Theory Division for warm hospitality as this work was being finished. He also thanks the organizers of the \textit{Large charge at Les Diablerets} meeting (as well as to SwissMAP) for a very stimulating environment where part of this work was carried out. D.R-G is partially supported by the Spanish government grant MINECO-16-FPA2015-63667-P. He also acknowledges support from the Principado de Asturias through the grant FC-GRUPIN-IDI/2018/000174. J.G.R. acknowledges financial support from projects  MINECO
grant PID2019-105614GB-C21, and  from the State Agency for Research of the Spanish Ministry of Science and Innovation through the “Unit of Excellence María de Maeztu 2020-2023” (CEX2019-000918-M).

\begin{appendix}

\section{Some properties of the one-point function}\label{1pointappendix}

In this appendix we will review some basic aspects about 1-point functions in the presence of a defect. As we have argued in the main text, the violation of conformal invariance in the bulk due to the non-zero $\beta$ function of the bulk coupling only enters at the two-loop order. Thus, to one-loop order we effectively have a defect CFT (dCFT). 

\medskip

On general grounds, in a dCFT,  we expect a bulk-defect OPE, which  allows us to write bulk operators $O$ in terms of defect operators $\mathcal{O}$ (see \textit{e.g.} \cite{Billo:2016cpy}),
\begin{equation}
O = \sum_{\mathcal{O}\in {\rm defect}}\, c_{O\mathcal{O}}\,\frac{\mathcal{O}}{|\vec{x}_T|^{\Delta_{O}-\Delta_{\mathcal{O}}}}\,.
\end{equation}
We recall that $d_T=d-d_{\rm wv}$, so $d_T=3-\epsilon $ in the 4d theory and $d_T=4-\epsilon $ in the 6d theory.
Since the defect is at a fixed point, only the identity operator can take a VEV. As a consequence, it immediately follows upon taking a VEV that the bulk 1-point function is

\begin{equation}
\label{1point}
\langle O(\vec{x}_T)\rangle=\frac{c_O}{|\vec{x}_T|^{\Delta_O}}\,.
\end{equation}

This formula can be explicitly verified by direct computation. Consider the computation of  the 1-point function of the bulk field from the path integral. In the saddle-point approximation, this boils down  to evaluating the integrand in the saddle-point solution as

\begin{equation}
\langle \phi_i\rangle=\int \mathcal{D}\phi\,\phi_i\,e^{-S}=\phi_i\Big|_{\rm saddle}=\phi_i^{(0)}+\phi_i^{(1)}\,,
\end{equation}
where $\phi_i^{(0)},\,\phi_i^{(1)} $ stand for the saddle-point solutions in (2.9). 
As an example, let us focus on the leading term,

\begin{equation}
\langle \phi_i\rangle =-\nu_i\,\int d^dz\,G(x-z)\,\delta_T(z)\,.
\end{equation}
Inserting the propagator explicitly, we get

\begin{equation}
\langle \phi_i\rangle =-\nu_i\,\int d^dz\,\int \frac{d^dp}{(2\pi)^d}\,\frac{e^{ip(x-z)}}{p^2}\,\delta_T(z)\,.
\end{equation}
Therefore

\begin{equation}
\langle \phi_i\rangle =-\nu_i \int \frac{d\vec{p}_{T}}{(2\pi)^{d_{T}}}\,\frac{e^{i\vec{p}_{T}\cdot\vec{x}_{T}}  }{|\vec{p}_{T}|^2}\,.
\end{equation}
Next, performing the Fourier transform, we obtain

\begin{equation}
\langle \phi\rangle \sim  -\frac{\nu}{|x_T|^{d_T-2}} = \begin{cases}   -\frac{\nu}{|x_T|}\,, \quad \   d=4,\,d_T=3\ ,\\  -\frac{\nu}{|x_T|^2}\, ,\quad d=6,\,d_T=4\ ; \end{cases}
\end{equation}
which precisely reproduces \eqref{1point}.

It is also easy to make contact with the results in the main text. Indeed, consider integrating the 1-point function above against a delta function in the transverse space. A simple computation shows that 

\begin{equation}
\langle \int d^dx\, \phi\,\delta_T(\vec{x})\rangle =-\nu\, \mathcal{V}\, \int \frac{d\vec{p}_{T}}{(2\pi)^{d_{T}}}\,
\frac{1}{|\vec{p}_{T}|^{2}}\,.
\end{equation}
where $\mathcal{V}=\int dx_{||}$ is the regularized defect volume.
This  reproduces the leading term in (3.3). 

The next contributions in the $\epsilon $ expansion can be computed
systematically, similarly as we did in sections 2--4 (see \cite{Rodriguez-Gomez:2022gbz}). In particular, in $d=4$,
for a dCFT (that is, a theory sitting in bulk and defect fixed points)
one still obtains \eqref{1point} with $\Delta_\phi$ 
given by 
an expansion of the form (see {\it e.g.} \cite{Allais:2014fqa})
$\Delta_\phi=1- \frac12 \epsilon+ b \epsilon^2+...$. 
In momentum space, this is
\begin{equation}
\label{yya}
\langle \tilde \phi(p)\rangle \sim  -\frac{{\cal N}}{|p_T|^{2-\epsilon/2-b\epsilon^2+...}} \ .
\end{equation}
To $O(\epsilon)$, this is indeed the form of the one-point function \eqref{granfinale}.
However, in the present case, as described in section 4, at the two-loop order the effect of the non-zero bulk $\beta$ function kicks in. As a consequence, the one-point function
\eqref{granfinale} to $O(\epsilon^2)$ does not have the form
\eqref{yya}. 
While there is still a bulk-defect OPE, the $c_{O\mathcal{O}}$ now become functions of the dimensionless combination $\mu\,|\vec{x}_T|$. Thus, instead of \eqref{1point}, one now finds that

\begin{equation}
\label{1point-non-CFT}
\langle O(\vec{x}_T)\rangle=\frac{c_O}{|\vec{x}_T|^{\Delta_O}}\,f(\mu\,|\vec{x}_T|)\,.
\end{equation}
As shown in section 4.1, the form of $f$ is constrained by the fact that the 1-point function is subject to a Callan-Symanzik equation. We have shown that this constrained form of $f$  agrees
with the $f$ obtained independently by direct calculation in terms of Feynman diagrams.

\section{Integrals in $d=4-\epsilon$}

In this subsection, we compute the $\mathcal{I}_1$ and $\mathcal{I}_2$ integrals in $d=4-\epsilon $, where $n=4$.

\subsection{The $\mathcal{I}_1$ integral}

When $n=4$, the integral $\mathcal{I}_1$ is given by 

\begin{equation}
\mathcal{I}_1=\int  \frac{d\vec{k}_T^1}{(2\pi)^{d_T}} \int  \frac{d\vec{k}_T^{2}}{(2\pi)^{d_T}}   \frac{1}{(\vec{k}_T^1)^2\,(\vec{k}^{2}_T)^2 \, (\vec{p}_T-\vec{k}_T^1-\vec{k}_T^{2})^2} \, .
\end{equation}
This integral has been recently computed in eq.(114) in \cite{Rodriguez-Gomez:2022gbz}, using the general formulas of \cite{Grozin:2003ak}. One finds

\begin{equation}
\int\frac{d\vec{k}_T^1}{(2\pi)^{d_T}}\,\int\frac{d\vec{k}_T^2}{(2\pi)^{d_T}}\, \frac{1}{(\vec{k}_T^1)^2\,(\vec{k}_T^2)^2\,(\vec{k}_T^3-\vec{k}_T^1-\vec{k}_T^2)^2}=\frac{\pi^{d-1}}{(2\pi)^{2d-2}}\,G(1,1)\,G(1,\frac{5-d}{2})\,|\vec{p}_T|^{2d-8}\,.
\end{equation}
where
\begin{equation}
    G(n_1,n_2)\equiv \frac{\Gamma(n_1+n_2-d_T/2)\Gamma(d_T/2-n_1)\Gamma(d_T/2-n_2)}{\Gamma(n_1)\Gamma(n_2)\Gamma(d_T-n_1-n_2)}\ ,\qquad d_T=d-d_{\rm wv}\ ,
\end{equation}
with $d_{\rm wv}=1$ in the 4d theory and $d_{\rm wv}=2$ in the 6d theory.
Expanding at small $\epsilon $, we obtain

\begin{equation}
\label{aacu}
\mathcal{I}_1=\frac{\Omega}{\epsilon}-2\,\Omega\,\log|\vec{p}_T|+A\epsilon+2\Omega\epsilon\,(\log|\vec{p}_T|)^2\, ,\qquad A\equiv \frac{9\,\Omega}{2}-\frac{7}{768}\ .
\end{equation}
We have omitted the dependence on $\mu$, which can be restored by $|\vec{p}_T|\to |\vec{p}_T|/\mu$.
 We have also absorbed a  numerical constant in the definition of $\mu$ by the rescaling $\mu^{-1}\,|\vec{p}_T|\rightarrow \mu^{-1}\,|\vec{p}_T|\,e^{\frac12 (3-\gamma_E+\log(4\pi))}$.

\subsection{The $\mathcal{I}_2$ integral}

For $\mathcal{I}_2$ we have

\begin{eqnarray}
\mathcal{I}_2 &=&\int\frac{d\vec{k}_T^1}{(2\pi)^{d_T}}\,\int\frac{d\vec{k}_T^2}{(2\pi)^{d_T}}\,\int\frac{d\vec{k}_T^3}{(2\pi)^{d_T}}\,\int\frac{d\vec{k}_T^4}{(2\pi)^{d_T}} 
\nonumber\\
&&\frac{1}{(\vec{k}_T^1)^2\,(\vec{k}_T^2)^2\,(\vec{k}_T^3)^2\,(\vec{k}_T^4)^2\,(\vec{k}_T^3-\vec{k}_T^1-\vec{k}_T^2)^2\,(\vec{p}_T-\vec{k}_T^3-\vec{k}_T^4)^2}
\,.
\end{eqnarray}
This can be written  as

\begin{eqnarray}
&& \mathcal{I}_2=\int\frac{d\vec{k}_T^3}{(2\pi)^{d_T}}\, \frac{1}{(\vec{k}_T^3)^2}\,  \nonumber \\ && \Big[\int\frac{d\vec{k}_T^1}{(2\pi)^{d_T}}\,\int\frac{d\vec{k}_T^2}{(2\pi)^{d_T}}\, \frac{1}{(\vec{k}_T^1)^2\,(\vec{k}_T^2)^2\,(\vec{k}_T^3-\vec{k}_T^1-\vec{k}_T^2)^2}\Big]\, \Big[\int\frac{d\vec{k}_T^4}{(2\pi)^{d_T}} \frac{1}{(\vec{k}_T^4)^2\,(\vec{p}_T-\vec{k}_T^3-\vec{k}_T^4)^2}\Big]
\,. \nonumber
\end{eqnarray}
From eq.(114) in \cite{Rodriguez-Gomez:2022gbz} (see also \cite{Grozin:2003ak})

\begin{equation}
\int\frac{d\vec{k}_T^1}{(2\pi)^{d_T}}\,\int\frac{d\vec{k}_T^2}{(2\pi)^{d_T}}\, \frac{1}{(\vec{k}_T^1)^2\,(\vec{k}_T^2)^2\,(\vec{k}_T^3-\vec{k}_T^1-\vec{k}_T^2)^2}=\frac{\pi^{d-1}}{(2\pi)^{2d-2}}\,|\vec{k}_T^3|^{2(d-4)}\, G(1,1)\,G(1,\frac{5-d}{2})\,.
\end{equation}
In turn the second integral can be read off from eq.(123) in \cite{Rodriguez-Gomez:2022gbz}, 

\begin{equation}
\int\frac{d\vec{k}_T^4}{(2\pi)^{d_T}} \frac{1}{(\vec{k}_T^4)^2\,(\vec{p}_T-\vec{k}_T^3-\vec{k}_T^4)^2}=\frac{2^{5-2d}\,\pi^{2-\frac{d}{2}}}{\cos\big(\frac{d\pi}{2}\big)\,\Gamma\big(\frac{d-2}{2}\big)} \frac{1}{(\vec{p}_T-\vec{k}_T^3)^{5-d}}\,.
\end{equation}
Thus we need to compute

\begin{equation}
\mathcal{I}_2=\frac{ 2^{7-4d}\,\pi^{3-\frac{3d}{2}}}{\cos\big(\frac{d\pi}{2}\big)\,\Gamma\big(\frac{d-2}{2}\big)}\, G(1,1)\,G(1,\frac{5-d}{2})\,\int\frac{d\vec{k}_T}{(2\pi)^{d_T}}\, \frac{1}{(\vec{k}_T)^{2\,(5-d)}\,(\vec{k}_T-\vec{p})^{5-d}}\, 
\end{equation}
Using \cite{Grozin:2003ak} 

\begin{equation}
\int\frac{d\vec{k}_T}{(2\pi)^{d_T}}\, \frac{1}{(\vec{k}_T)^{2\,(5-d)}\,(\vec{k}_T-\vec{p})^{5-d}}=\frac{\pi^{\frac{d-1}{2}}}{(2\pi)^{d-1}}\,|\vec{p}_T|^{4\,(d-4)} \,G(5-d,\frac{5-d}{2})\, 
\end{equation}
Combining the above expressions, we get

\begin{equation}
\mathcal{I}_2=\frac{2^{8-5d}\,\pi^{\frac{7}{2}-2d}}{\cos\big(\frac{d\pi}{2}\big)\,\Gamma\big(\frac{d-2}{2}\big)}\,G(1,1)\,G(1,\frac{5-d}{2})\,G(5-d,\frac{5-d}{2})\,|\vec{p}_T|^{4(d-4)}\,.
\end{equation}
Expanding in powers of $\epsilon$, we find 

\begin{equation}
\mathcal{I}_2=\frac{\Omega^2}{2\,\epsilon^2}+\frac{\Omega^2\,(1-2\log|\vec{p}_T|)}{\epsilon}+\Big(C -4\,\Omega^2\,\log|\vec{p}_T|+4\,\Omega^2\,(\log|\vec{p}_T|)^2\Big)\,.
\end{equation}
where $C=\frac{19\,\Omega^2}{2}-\frac{11\,\Omega}{768}$
and 
 we have rescaled $\mu$ by the same numerical constant as in \eqref{aacu}. 

\section{Integrals in $d=6-\epsilon$}

We now compute the $\mathcal{I}_1$ and $\mathcal{I}_2$ integrals near $d=6$, where $n=3$.

\subsection{The $\mathcal{I}_1$ integral}

For $\mathcal{I}_1$ we have

\begin{equation}
\mathcal{I}_1=\int  \frac{d\vec{k}_T}{(2\pi)^{d_T}}  \frac{1}{ \vec{k}_T^2\,(\vec{p}_T-\vec{k}_T)^2} \, .
\end{equation}
We can directly borrow (123) in \cite{Rodriguez-Gomez:2022gbz}. In $6-\epsilon$ dimensions,

\begin{equation}
\label{bbcu}
\mathcal{I}_1=\frac{\Omega}{\epsilon}-\Omega\,\log|\vec{p}_T|+A\epsilon+\epsilon\,\frac{\Omega}{2}\,(\log|\vec{p}_T|)^2 \, ,\qquad A\equiv \frac{\Omega}{2}-\frac{1}{384}\ .
\end{equation}
Again, we have omitted the dependence on $\mu$, which is restored by $|\vec{p}_T|\to |\vec{p}_T|/\mu$, and
  absorbed a  numerical constant in the definition of $\mu$ by $\mu^{-1}\,|\vec{p}_T|\rightarrow \mu^{-1}\,|\vec{p}_T|\,e^{\frac12 (2-\gamma_E+\log(4\pi))}$.

\subsection{The $\mathcal{I}_2$ integral}

In the 6d model, the  $\mathcal{I}_2$ integral is given by

\begin{equation}
\mathcal{I}_2= \int  \frac{d\vec{k}_T^1}{(2\pi)^{d_T}}  \int \frac{d\vec{k}_T^{2}}{(2\pi)^{d_T}} \frac{1}{(\vec{k}_T^1)^2 \, (\vec{k}_T^{2})^2\, (\vec{k}_T^{2}-\vec{k}_T^1)^2\,(\vec{k}_T^2-\vec{p}_T)^2} 
\,.
\end{equation}
This integral is a particular case of the integral computed in eq. (2.8) in \cite{Grozin:2003ak}, with $n_1=0$, $n_2=1$, $n_3=1$, $n_4=1$, $n_5=1$. This gives

\begin{equation}
\mathcal{I}_2=\frac{\pi^{d-2}}{(2\pi)^{2(d-2)}}\,|\vec{p}_T|^{2d-12}\,G(1,1)\,G(1,\frac{8-d}{2})\,.
\end{equation}
Expanding in powers of $\epsilon$, we obtain

\begin{equation}
\mathcal{I}_2=\frac{\Omega^2}{2\epsilon^2}+\frac{\Omega^2\,(1-4\,\log|\vec{p}_T|)}{4\epsilon}+\Big(C-\frac12 \Omega^2\,\log|\vec{p}_T|+\Omega^2\,(\log|\vec{p}_T|)^2\Big)\,.
\end{equation}
where, in the 6d model, we define $C\equiv \frac{7\Omega^2}{8}-\frac{\Omega}{384}$. We have rescaled $\mu $
by the same numerical constant as in \eqref{bbcu}.

\end{appendix}

\end{document}